\documentclass[traditabstract,a4paper]{aa}

\pdfoutput=1

\usepackage{newtxtext,newtxmath}
\usepackage{graphicx}
\usepackage{natbib}
\usepackage{rotating}

\bibpunct{(}{)}{;}{a}{}{,} % to follow the A&A style

\hypersetup{pdfauthor={E. O'Sullivan},
            pdftitle={Cold gas in a complete sample of group-dominant early-type galaxies},
            pdfkeywords={Galaxies: elliptical and lenticular, cD; Galaxies: groups: general; Galaxies: star formation; Radio lines: galaxies},
            bookmarksnumbered=true}

\newcommand{\arcm}{\hbox{$^\prime$}}

\newcommand{\degree}{\hbox{$^\circ$}}

\newcommand{\chandra}{\emph{Chandra}}
\newcommand{\xmm}{\emph{XMM-Newton}}

\newcommand{\arcs}{\mbox{\arcm\arcm}}

\newcommand{\Lsol}{\ensuremath{L_{\odot}}}
\newcommand{\Msol}{\ensuremath{M_{\odot}}}
\newcommand{\Msolpyr}{\ensuremath{M_{\odot}~yr^{-1}}}

\newcommand{\K}{\ensuremath{\mbox{~K}}}
\newcommand{\s}{\ensuremath{\mbox{~s}}}
\newcommand{\ps}{\ensuremath{\s^{-1}}}
\newcommand{\cm}{\ensuremath{\mbox{~cm}}}
\newcommand{\pcmsq}{\ensuremath{\cm^{-2}}}
\newcommand{\km}{\ensuremath{\mbox{~km}}}

\newcommand{\erg}{\ensuremath{\mbox{~erg}}}
\newcommand{\ergps}{\ensuremath{\erg \ps}}

\newcommand{\kmps}{\ensuremath{\km \ps}}
\newcommand{\Kkmps}{\ensuremath{\K .\km \ps}}

\newcommand{\gtsim}{\,\rlap{\raise 0.5ex\hbox{$>$}}{\lower 1.0ex\hbox{$\sim$}}\,} 

\newcommand{\Hi}{H\textsc{i}}

\newcommand{\Tmb}{\ensuremath{\mathrm{T}_\mathrm{mb}}}
\begin{document}

\title{Cold gas in a complete sample of group-dominant early-type galaxies
\thanks{Based on observations carried out with the IRAM 30m telescope and the Atacama Pathfinder Experiment (APEX). IRAM is supported by INSU/CNRS (France), MPG (Germany), and IGN (Spain). APEX is a collaboration between the Max-Planck-Institut fur Radioastronomie, the European Southern Observatory, and the Onsala Space Observatory.}
}

\author{E. O'Sullivan \inst{1}
\and
F. Combes \inst{2,3}
\and
P. Salom\'e \inst{2}
\and
L.~P. David \inst{1}
\and
A. Babul \inst{4}
\and
J.~M. Vrtilek \inst{1}
\and
J. Lim \inst{5,6}
\and
V. Olivares \inst{2}
\and \\
S. Raychaudhury \inst{7,8}
\and
G. Schellenberger \inst{1}
           }
\offprints{E. O'Sullivan}
\institute{Harvard-Smithsonian Center for Astrophysics, 60 Garden Street,
Cambridge, MA 02138, USA\\
\email{eosullivan@cfa.harvard.edu}
 \and
LERMA, Observatoire de Paris, CNRS, PSL Univ., Sorbonne Univ., 75014 Paris, France
 \and
Coll\`{e}ge de France, 11 place Marcelin Berthelot, 75005 Paris, France
\and
Department of Physics and Astronomy, University of Victoria, Victoria, BC, V8W 2Y2, Canada
\and
Department of Physics, The University of Hong Kong, Pokfulam Road, Hong Kong
\and
 Laboratory for Space Research, Faculty of Science, The University of Hong Kong, Pokfulam Road, Hong Kong
\and
Inter-University Centre for Astronomy and Astrophysics, Pune 411007, India
\and
Department of Physics, Presidency University, 86/1 College Street, 700073 Kolkata, India
              }

\date{Received 5 June 2018 / Accepted 23 July 2018}

\titlerunning{CO in group-dominant ellipticals}
\authorrunning{E. O'Sullivan et al.}

%300 word maximum
\abstract{We present IRAM 30m and APEX telescope observations of CO(1-0) and CO(2-1) lines in 36 group-dominant early-type galaxies, completing our molecular gas survey of dominant galaxies in the Complete Local-volume Groups Sample. We detect CO emission in 12 of the galaxies at $>$4$\sigma$ significance, with molecular gas masses in the range $\sim$0.01-6$\times$10$^8$\Msol, as well as CO in absorption in the non-dominant group member galaxy NGC~5354. In total 21 of the 53 CLoGS dominant galaxies are detected in CO and we confirm our previous findings that they have low star formation rates (0.01-1\Msolpyr) but short depletion times ($<$1~Gyr) implying rapid replenishment of their gas reservoirs. Comparing molecular gas mass with radio luminosity, we find that a much higher fraction of our group-dominant galaxies (60$\pm$16\%) are AGN-dominated than is the case for the general population of ellipticals, but that there is no clear connection between radio luminosity and the molecular gas mass. Using data from the literature, we find that at least 27 of the 53 CLoGS dominant galaxies contain \Hi, comparable to the fraction of nearby non-cluster early type galaxies detected in \Hi\ and significantly higher that the fraction in the Virgo cluster. We see no correlation between the presence of an X-ray detected intra-group medium and molecular gas in the dominant galaxy, but find that the \Hi-richest galaxies are located in X-ray faint groups. Morphological data from the literature suggests the cold gas component most commonly takes the form of a disk, but many systems show evidence of galaxy-galaxy interactions, indicating that they may have acquired their gas through stripping or mergers. We provide improved molecular gas mass estimates for two galaxies previously identified as being in the centres of cooling flows, NGC~4636 and NGC~5846, and find that they are relatively molecular gas poor compared to our other detected systems.}

\maketitle

%---------------------------------------------------------------

\section{Introduction}
\label{sec:intro}

Early-type galaxies, and particularly those which occupy the centres of galaxy groups and clusters, are often treated as an homogeneous class, with old red stellar populations and little or no cold gas content or star formation. In recent years, several lines of research \citep[e.g.,][]{Bildfelletal08,Loubseretal16} have highlighted inaccuracies in this picture, in particular that elliptical and lenticular galaxies across all environments can possess a cold gas component suitable for fuelling star formation or an active galactic nucleus (AGN). Studies of early-type galaxies in the local universe have shown that many contain significant reservoirs of both atomic and molecular gas \citep{Morgantietal06,Combesetal07,Oosterlooetal10}, with statistical studies showing that $\sim$20\% are detected in CO \citep{Youngetal11} and $\sim$40\% in \Hi, although this fraction falls to $\sim$10\% in the Virgo cluster \citep{Serraetal12}. While these reservoirs are smaller than those observed in spiral galaxies, they can still be substantial. Masses of $\sim$10$^8$\Msol\ of atomic and $\sim$10$^7$\Msol\ of molecular gas are not uncommon. The morphology and kinematics of the gas can provide clues to its origin. While gas disks and rings are common \citep{Crockeretal11,Chungetal12,Davisetal13,Youngetal18}, some of these are kinematically offset from the stellar population \citep[e.g.,][]{Crockeretal08}, and a significant number of early-type galaxies host disturbed structures likely arising from tidal interactions \citep{Youngetal08,Alataloetal13,LuceroYoung13}.

Some cluster-dominant ellipticals have also been shown to possess cold gas reservoirs containing up to 10$^{11}$\Msol\ of molecular gas \citep{Edgeetal01,SalomeCombes03,Pulidoetal18}. Imaging studies have shown that molecular gas in brightest cluster galaxies (BCGs) tends to be located in filamentary structures associated with H$\alpha$ nebulae \citep{Salomeetal06,McDonaldetal11,Salomeetal11,Limetal12,Russelletal16,Vantyghemetal16,Russelletal17a,Russelletal17b,Vantyghemetal17}. Molecular gas is only observed in clusters with low central entropies or cooling times shorter than 1~Gyr \citep{Pulidoetal18} strongly suggesting that it is the product of cooling from the hot, X-ray emitting intra-cluster medium (ICM). The molecular gas, which typically has low velocity relative to the BCG, is of particular importance as a component of the AGN feedback cycle in clusters. Numerical modelling suggests that rapid cooling in thermally unstable regions of the ICM can produce the observed cold, dense clouds \citep[e.g.,][]{Sharmaetal12,McCourtetal12,Gasparietal12,LiBryan14a,LiBryan14b,Prasadetal15,Prasadetal17}, though there is still some debate as to how this condensation is triggered \citep{McNamaraetal16,Prasadetal17,Gasparietal18,Prasadetal18}.  

Galaxy groups occupy the interesting intermediate scale between individual galaxies and massive galaxy clusters. Many are observed to possess a hot ($\sim$10$^7$~K) intra-group medium (IGM) and $\sim$60\% of these have central temperature declines indicative of rapid radiative cooling \citep{OSullivanetal17}. Some X-ray bright groups show striking similarities to galaxy clusters, with short central cooling times, H$\alpha$ nebulae and molecular gas clouds in their central ellipticals \citep[e.g.,][]{Davidetal14,Davidetal17,Temietal17} and evidence of AGN feedback in the form of jet-inflated cavities in the IGM \citep[e.g.,][]{OSullivanetal05b,Baldietal09b,Gastaldelloetal09,Machaceketal11,Davidetal11}. However, groups are a very diverse class, and their low velocity dispersions and small galaxy separations make interactions and mergers a much more important factor than in clusters. The group environment is known to have an impact on member galaxies, with the cold gas content of individual late-type galaxies and of the group as a whole declining with increasing mass \citep[e.g.,][]{Kilbornetal09,Desjardinsetal14,Odekonetal16}. Tidal stripping of cold gas can produce intergalactic gas clouds and filaments \citep[e.g.,][]{Appletonetal90,Williamsetal02,Durbalaetal08,Oosterlooetal18}, or even a cold IGM \citep{VerdesMontenegroetal01,Johnsonetal07,Konstantopoulosetal10,Borthakuretal10}, and ram-pressure effects, while not so common, have been observed in some cases \citep[e.g.,][]{Rasmussenetal06a}. 

An understanding of the cold gas content of group dominant galaxies is therefore important for several reasons, including its role in the feedback cycle, its ability to fuel star formation in these galaxies that were once thought to be "red and dead", and as a tracer of the galaxy interactions which transport and transform cold gas within groups.

In this paper we present 33 new IRAM 30m and APEX observations of group-dominant early-type galaxies, as well as results from 3 archival APEX observations. In combination with our earlier study \citep[hereafter referred to as paper I]{OSullivanetal15}, these provide complete coverage of the brightest group early-type (BGE) galaxies of the 53-group Complete Local-Volume Group Sample, as well as deep integrations on 2 additional X-ray bright cool core groups chosen as likely examples of IGM cooling driving AGN feedback. Our goal with this study is to determine the fraction of group-dominant galaxies which contain significant molecular gas reservoirs, to examine the origin of this material, and to study its relationship to group properties and nuclear activity. 

Our sample is described in Section~\ref{sec:samp}, and our observations in Section~\ref{sec:obs}. We describe our results in Section~\ref{sec:results}, including the detection fraction, connection to star formation and AGN, and relationship with group properties, and discuss the possible origin of the cold gas in Section~\ref{sec:disc}. We give our conclusions in Section~\ref{sec:conc}. Distances are based on the Virgocentric-flow corrected recession velocity drawn from HyperLEDA and an assumed H$_0$=70\kmps~Mpc$^{-1}$, except in cases where a redshift-independent distance is available from the surface brightness fluctuation (SBF) catalogue of \citet{Tonryetal01}. These SBF distances are corrected to match the Cepheid zero point of \citet{Freedmanetal01} as described in \citet{Jensenetal03}.

\section{Sample}
\label{sec:samp}

The Complete Local-Volume Groups Sample \citep[CLoGS,][]{OSullivanetal17} consists of 53 groups in the local universe (Distance $<$ 80~Mpc). CLoGS aims to provide a statistically complete, representative sample of nearby groups whose properties can be studied across a range of wavebands. A combined optical and X-ray selection approach is used to avoid the biases which can effect either selection individually. The groups were chosen from the Lyon Galaxy Group catalogue of \citep{Garcia93}, which used a combination of friends-of-friends and hierarchical clustering algorithms to select groups from an early version of the Lyon Extragalactic Data Archive\footnote{http://leda.univ-lyon1.fr}. X-ray observations were then to be used to determine whether each group hosts a hot intra-group medium (IGM). The ability to heat and retain an IGM demonstrates that the group is a virialized system, and as the groups are nearby, the temperature and density structure of the IGM can be resolved, providing insights into the dynamical and thermal state of the group. The sample is restricted to declination $>$ -30\degree\ to ensure visibility from the Giant Metrewave Radio Telescope (GMRT) and Very Large Array (VLA), since multi-band radio observations can provide complementary information on AGN and star formation in the group member galaxies. This is of particular interest for the group-dominant early-type galaxies, since feedback heating from their AGN is expected to govern the thermal balance of the group, preventing excessive cooling and star formation. To date, $\sim$70\%  of the CLoGS groups have been observed in the X-ray by \xmm\ and/or \chandra, and observations of the remaining systems have been approved. GMRT 235 and 610~MHz radio continuum observations of all groups have been completed. A detailed description of sample selection and results for the 26-group high-richness subsample are presented in \citet{OSullivanetal17}.

As a complete sample of groups in the local volume with high-quality X-ray and radio observations, CLoGS provides an ideal sample with which to study the prevalence and origins of molecular gas in group-dominant galaxies. In \citet{OSullivanetal15} we presented CO measurements for a preliminary subset of 23 CLoGS group-dominant galaxies, observed with the IRAM 30m telescope. In this paper we present data for the remainder of the sample, based on new IRAM 30m observation for northern targets, and a combination of new and archival APEX observations for the southern groups. We also reobserved two galaxies of particular interest included in paper~I, \object{NGC~4261} and \object{NGC~5353}. The basic properties of the 32 galaxies for which we present CO measurements are shown in Table~\ref{tab:basic}.

In addition to completing the CLoGS sample, we also targeted a small number of group dominant elliptical galaxies for deeper observations, including \object{NGC~4636} and \object{NGC~5813} \citep[not part of CLoGS,][explains why they were not selected]{OSullivanetal17}, \object{NGC~5846} and \object{NGC~7619}. These systems were selected to be in highly X-ray luminous, cool-core groups with multiple indicators of cooling. Three of the four galaxies are detected in [C\textsc{ii}] emission \citep{Werneretal14}, a strong indicator of the presence of molecular gas thought to arise from photo-dissociation regions surrounding molecular clouds. All four are detected in H$\alpha$ \citep{Werneretal14,Macchettoetal96}, and all have radio-detected AGN, indicating ongoing accretion in the central engine. They also have low measured values of the thermal instability criterion (cooling time over free fall time, t$_c$/t$_{ff}$$\sim$10-25) in their hot IGM. 

\begin{table*}
\caption{Basic data for the newly observed or processed group-dominant galaxies}
\label{tab:basic}
\centering
\begingroup % keep narrower columns for this table only
\setlength{\tabcolsep}{4pt} % narrower columns
\begin{tabular}{lccccccccccc}
\hline\hline\\[-3mm]
Galaxy  & z  & D$_{L}$ & Beam$^a$ & log L$_{\rm B}^b$  & D$_{25}^b$ & log M$_{\rm dust}^d$ & log M(HI)$^c$ 
& log L$_{\rm FIR}^d$  & log M$_{*}^e$  & M(H$_2$)$^f$ &  F$_{1.4GHz}^g$ \\
        &    &  [Mpc]    & [kpc]     & [\Lsol]      &[kpc]       &[\Lsol]      &[\Msol]
 &[\Lsol]     &[\Msol]  & [10$^8$\Msol]&  [mJy] \\[+1mm]
\hline\\[-3mm]
\multicolumn{11}{l}{\textit{IRAM 30m CLoGS survey observations}}\\
NGC~128     & 0.014146 & 58.54 & 6.22 & 10.66 & 53.73 & 6.48 & 8.65    &  9.84$\pm$0.03 & 11.09 & 1.76$\pm$0.18$^h$ & 1.5 \\
NGC~252     & 0.016471 & 71.65 & 7.63 & 10.56 & 33.72 & 6.97 & 9.56    &  9.84$\pm$0.05 & 11.06 & 6.29$\pm$0.79    & 2.5 \\
NGC~410     & 0.017659 & 76.67 & 8.17 & 10.95 & 52.77 & -    & -       & -              & 11.37 & $<$1.17          & 5.8 \\
NGC~584     & 0.006011 & 18.71 & 1.97 & 10.20 & 20.60 & 6.06 & 8.08    &  8.15$\pm$0.06 & 10.62 & $<$0.11          & 0.6 \\
NGC~924     & 0.014880 & 63.61 & 6.76 & 10.30 & 37.61 & 7.30 & 9.96    &  9.50$\pm$0.23 & 10.76 & 0.52$\pm$0.10    & 0.9 \\
NGC~978     & 0.015794 & 68.66 & 7.30 & 10.56 & 36.60 & 5.49 & -       &  8.96$\pm$0.08 & 11.08 & $<$0.70          & 0.7 \\
NGC~1106    & 0.014467 & 63.26 & 6.72 & 10.33 & 33.56 & 6.19 & -       & 10.03$\pm$0.06 & 10.87 & 6.81$\pm$0.33    & 132 \\
NGC~1453    & 0.012962 & 53.54 & 5.68 & 10.61 & 38.32 & 7.15 & 9.01    &  9.10$\pm$0.06 & 11.16 & $<$0.78          & 28 \\
NGC~1550    & 0.012389 & 52.13 & 5.53 & 10.36 & 26.17 & -    & -       & -              & 10.88 & $<$0.47          & 17 \\
NGC~2563    & 0.014944 & 64.39 & 6.84 & 10.47 & 37.89 & -    & $<$8.73 & -              & 11.02 & $<$0.93          & 0.7 \\
NGC~2911    & 0.010617 & 45.84 & 4.85 & 10.43 & 46.13 & 5.13 & 9.31    &  8.83$\pm$0.07 & 10.82 & 2.66$\pm$0.31    & 56 \\
NGC~3325    & 0.018873 & 79.67 & 8.50 & 10.40 & 29.24 & -    & 9.18    &  4.10$\pm$0.01 & 10.88 & $<$1.57          & - \\
NGC~4008    & 0.012075 & 53.96 & 5.72 & 10.47 & 32.57 & 4.62 & $<$8.49 &  8.87$\pm$0.20 & 10.90 & $<$0.73          & 11 \\
NGC~4169    & 0.012622 & 56.67 & 6.01 & 10.44 & 27.04 & 7.48 & 10.03   & 10.46$\pm$0.01 & 10.89 & 1.44$\pm$0.34    & 1.1\\
NGC~4261    & 0.007378 & 29.38 & 3.10 & 10.65 & 36.29 & 5.04 & $<$8.69 &  8.30$\pm$0.11 & 11.05 & $<$0.37          & 19700 \\
NGC~4956    & 0.015844 & 70.90 & 7.54 & 10.58 & 32.16 & 5.87 & 9.05    &  9.54$\pm$0.09 & 10.91 & $<$0.75          & - \\
NGC~5353    & 0.007755 & 35.43 & 3.74 & 10.49 & 24.44 & 3.91 & 9.72    &  8.74$\pm$0.04 & 11.05 & 1.90$\pm$0.25    & 40 \\
%NGC~5354    & 0.008603 & 35.43 & 3.74 & 10.35 & 31.20 & 6.35 & 9.75 & 9.27$\pm$0.04 & 10.62 & - & 8.4 \\
NGC~5444    & 0.013169 & 60.46 & 6.42 & 10.64 & 43.67 & -    & $<$8.79 & -             & 11.04 & $<$0.64          & 660 \\
NGC~5490    & 0.016195 & 71.94 & 7.66 & 10.68 & 34.73 & -    & -       & -             & 11.15 & $<$0.64          & 1300 \\
NGC~5629    & 0.015004 & 67.73 & 7.20 & 10.58 & 33.61 & -    & -       & -             & 10.95 & $<$0.91          & 4.6 \\
NGC~6658    & 0.014243 & 63.97 & 6.79 & 10.66 & 27.84 & -    & -       & -             & 10.76 & $<$0.71          & - \\
\multicolumn{11}{l}{\textit{IRAM 30m deep observations}}\\
NGC~4636    & 0.003129 & 13.61 & 1.43 & 10.44 & 24.98 & 5.07 & 8.84$^i$    & 7.94$\pm$0.03 & 10.51 & 0.010$\pm$0.003  & 78 \\
NGC~5813    & 0.006525 & 29.92 & 3.15 & 10.53 & 36.28 & 4.55 & $<$7.94 & 8.09$\pm$0.06 & 10.89 & $<$0.11          & 15 \\
NGC~5846    & 0.005711 & 23.12 & 2.44 & 10.48 & 28.69 & 3.99 & 8.48$^i$ & 8.05$\pm$0.06 & 10.83 & 0.14$\pm$0.06    & 21 \\
NGC~7619    & 0.012549 & 54.30 & 5.76 & 10.82 & 39.68 & 6.98 & 7.41$^i$    & 9.17$\pm$0.12 & 11.21 & $<$0.29          & 20 \\
\multicolumn{11}{l}{\textit{APEX CLoGS survey observations}}\\
ESO~507-25  & 0.010788 & 45.21 & 6.00 & 10.45 & 34.67 & 6.50 & 10.50   &  9.46$\pm$0.05 & 10.95 & 4.23$\pm$0.56    & 24 \\
NGC~1779    & 0.011051 & 44.44 & 5.90 & 10.27 & 34.47 & 6.53 & 9.26    &  9.73$\pm$0.08 & 10.75 & 4.57$\pm$0.60    & 5.4 \\
NGC~2292    & 0.006791 & 29.57 & 3.91 & 10.40 & 37.29 & 6.84 & 9.35    &  9.26$\pm$0.03 & 10.79 & $<$0.47          & - \\
NGC~3923    & 0.005801 & 21.28 & 2.81 & 10.54 & 42.63 & 3.63 & $<$8.41 &  7.38$\pm$0.08 & 11.03 & $<$0.29          & 1.0 \\
NGC~5061    & 0.006945 & 28.26 & 3.73 & 10.61 & 30.82 & 2.65 & $<$8.66 &  7.40$\pm$0.10 & 10.96 & $<$0.43          & $<$1.0\\
NGC~5153    & 0.014413 & 60.54 & 8.06 & 10.59 & 43.03 & -    & -       & -              & 10.89 & $<$1.75          & - \\
NGC~5903    & 0.008556 & 31.48 & 4.17 & 10.30 & 27.46 & -    &  9.48   & -              & 10.73 & $<$0.58          & 320 \\
NGC~7377    & 0.011138 & 46.73 & 6.20 & 10.69 & 53.25 & 6.70 & -       &  9.47$\pm$0.05 & 11.04 & 4.74$\pm$0.44    & 3.1 \\
\multicolumn{11}{l}{\textit{APEX archival observations}}\\
NGC~1395    & 0.005727 & 22.39 & 2.95 & 10.65 & 31.39 & 5.64 & $<$8.45 &  8.12$\pm$0.08 & 10.91 & $<$0.27          & 1.1 \\
NGC~3078    & 0.008606 & 32.66 & 4.32 & 10.39 & 28.89 & -    & $<$8.78 & -              & 10.84 & $<$0.23          & 310 \\
NGC~5084    & 0.005741 & 24.09 & 3.18 & 10.31 & 69.59 & 5.74 & 10.18   &  8.87$\pm$0.05 & 10.91 & $<$0.16          & 46 \\
\hline
\end{tabular}
\endgroup
\tablefoot{
$^{a}$ FWHM CO(1-0) beam for IRAM observations, CO(2-1) beam for APEX observations.
$^{b}$ Computed from HYPERLEDA (http://leda.univ-lyon1.fr/).
$^{c}$ \Hi\ masses are drawn from NED (http://nedwww.ipac.caltech.edu/), the extragalactic distance database all-digital \Hi\ catalog \citep{Courtoisetal09}, \citet{SerraOosterloo10}, \citet{Serraetal12}, \citet{Appletonetal90}, \citet{Huchtmeier94}, \citet{Barnes99}, \citet{Serraetal08}, \citet{Chungetal12} and \citet{Haynesetal18}.
$^{d}$ The derivation of L$_{\rm FIR}$ and dust masses M$_{\rm dust}$ is described in Sec.~\ref{sec:SFR}.
$^{e}$ Stellar masses were obtained through SED fitting with SDSS and/or 2MASS fluxes.
$^{f}$ Molecular gas masses and upper limits either derived from our
  data or drawn from previous
  observations (see Section~\ref{sec:obs}). Where CO(1-0) and CO(2-1) estimates are available, the larger mass or upper limit is reported here.
$^{g}$ 1.4~GHz continuum fluxes are drawn from the NRAO VLA Sky Survey \citep[NVSS,][]{Condonetal98,Condonetal02}, the Faint Images of the Radio Sky at Twenty Centimeters survey \citep[FIRST,][]{Beckeretal95}, or from \citet{Brownetal11}\, except in the cases of NGC~924, NGC~978 and NGC~2563, where we extrapolate from the 610~MHz flux \citep{Kolokythasetal18} assuming a spectral index $\alpha$=0.8, and NGC~5903 where use the measured flux from \citet{OSullivanetal18}.
$^h$ mass estimated from offset pointing, see NGC~128-HI in Table~\ref{tab:det}. $^i$ conflicting H\textsc{i} measurements, see text for discussion.
}
\end{table*}

\section{Observations}
\label{sec:obs}

\subsection{IRAM 30m}

The IRAM 30m survey observations of the CLoGS group dominant early-type galaxies were performed on 2015 June 2-4 (program 192-14), 2017 June 2-5 and September 25-26 (program 064-17) in good weather conditions. Pointings and focus were done on strong quasars or planets every two hours. The EMIR receivers were tuned to the redshifted CO(1-0) and CO(2-1) frequencies for simultaneous observations in the E0 and E2 bands (which cover 73-117~GHz and 202-274~GHz respectively) in both horizontal and vertical polarisations. The FTS and WILMA backends were used at both 2.6mm and 1.3mm. We spent an average of $\sim$2 hours on each galaxy, and achieved noise levels \Tmb$\sim$1~mK (where \Tmb\ is the main beam temperature) smoothed over 50\kmps.

Observations of the four targets chosen for deep integrations were performed on 2017 June 8 (NGC~4636 and NGC~5846, program 063-17), 2017 December 29 and 2017 March 21-22 (NGC~5813 and NGC~7619, program 162-17). Between $\sim$6 and $\sim$19~hours were spent on each target, scaling with distance. Unfortunately, weather conditions during the December and January observations were poor, and we only achieved noise levels \Tmb=0.6mK and \Tmb=1.2mK, smoothed over 50\kmps. Setup was otherwise identical to the survey observations.

The results presented here are from the WILMA backend and cover a bandwidth of 3.7~GHz, equivalent to $\sim$9600\kmps\ at 2.6mm and $\sim$4800\kmps\ at 1.3mm. We used the GILDAS CLASS software to reduce the data. The instrumental resolution of 2~MHz was Hanning smoothed to 8~MHz (21\kmps) in order to increase the signal-to-noise. Spurious lines and bad spectra were removed. After a baseline subtraction, spectra were averaged and converted into main beam temperature. We used beam efficiencies of 0.78 and 0.59 and forward efficiencies of 0.95 and 0.92 at 2.6mm and 1.3mm respectively. 

\subsection{APEX}

Our APEX CO(2-1) observations of 8 southern BGEs were performed in service mode during 2017 July-August (project E-0100.a-0170A), using the APEX-1 receiver of the Swedish Heterodyne Facility Instrument (SHeFI). Observing conditions were generally good, with Precipitable Water Vapor (PWV) columns of 0.8-1.6~mm. Total times spent on-source were 27-87~min, achieving sensitivities of $\sim$0.9~mK, smoothed over 50\kmps. 

For three additional galaxies, archival APEX-1 CO(2-1) data were available in the ESO archive. NGC~1395 was observed on 2012 Apr 7 (project E-089.B-0866A, P.I. M.~Lehnert) in poor conditions (PWV=3-4~mm) for $\sim$28~min. NGC~3078 and NGC~5084 were observed on 2014 Jun 14 (project O-093.F-9309A, P.I. T.~Davis) under good conditions (PWV=0.6-0.9~mm) for $\sim$30 and $\sim$20~min respectively. We analysed these data and found sensitivities of 0.5-1.3~mK, smoothed over 50\kmps.

Reduction and analysis of the APEX data were carried out following the approach described above for the IRAM 30m. We used a beam efficiency of 0.75 and forward efficiency of 1.0 when converting from antenna to main beam temperatures.

\subsection{Common analysis steps}
When detected, spectral lines were fitted with a Gaussian profile, or if necessary (as was the case for NGC~1106) a two-Gaussian model. Line intensities were taken to be the area under the model profile, i.e.:

\begin{equation}
\mathrm{I_{CO} = 1.06\, \Tmb\, \Delta V },
\end{equation}

for a single Gaussian, where I$_{\rm CO}$ is the line intensity in \Kkmps, and $\Delta$V is the linewidth (FWHM) of the Gaussian. Where no line was detected, upper limits on the line intensities were derived for an upper limit on the peak of 3 times the rms and a fixed line width of 300\kmps (a typical value for early-type galaxies, and comparable to the mean linewidth in our observations) as follows:

\begin{equation}
\mathrm{I_{CO} < \frac{3 rms\, \Delta V}{\sqrt{N_{ch}}}},
\end{equation}

where N$_{\rm ch}$ is the number of channels in $\Delta$V. CO luminosities were computed following \citet{Solomonetal97}, with 

\begin{equation}
\mathrm{L^\prime_{CO} = 3.25\times 10^7\, \frac{S_{CO}\, D_L^2}{\nu_{obs}^2\, (1+z)^3}}, 
\end{equation}

where L$^\prime_{\rm CO}$ has units of \Kkmps~pc$^2$, $\nu_{\rm obs}$ is the observed frequency of the line, $\nu_{\rm rest}$/(1+z), and D$_L$ is the luminosity distance in Mpc. S$_{\rm CO}$ is the CO flux in Jy~km~s$^{-1}$, equal to I$_{\rm CO}$ multiplied by a telescope specific conversion factor $S/T_{mb}$=5~Jy~K$^{-1}$ for the IRAM 30m and $S/T_{mb}$=39~Jy~K$^{-1}$ for APEX. In order to estimate the corresponding molecular gas masses, the standard conversion factor for nearby quiescent galaxies like the Milky Way, $\alpha_{\rm CO}$=4.6 \citep{SolomonVandenbout05} was then used in 

\begin{equation}
\mathrm{M_{gas} = \alpha_{CO} L^\prime_{CO}} , 
\end{equation}

where M$_{\rm gas}$ is in \Msol\ and $\alpha_{\rm CO}$ has units of \Msol.(\Kkmps~pc$^2$)$^{-1}$. While detailed studies of the filamentary nebula at the center of the Perseus cluster suggest that radiatively cooled material may in some cases have a different conversion factor \citep{Limetal17}, the standard value provides a simple baseline for our relatively diverse sample.

As discussed in paper~I, the beam sizes of our observations ($\sim$2-8.5~kpc FWHM) are smaller than the stellar extent of the galaxies, and we cannot therefore be certain of capturing the total mass of CO in each system. Figure~\ref{fig:beams} shows the instrument beams overlaid on Digitized Sky Survey optical images of the BGEs. Interferometric mapping shows that CO is typically concentrated in galaxy cores, declining exponentially with radius \citep{YoungScoville91} and that very few early-type galaxies have CO extent $>$2~kpc \citep{Davisetal13}. Cool-core groups may be an exception, as their H$\alpha$ nebulae can extend several kpc \citep[e.g.,][]{Werneretal14}, indicating cooling on these scales. We discuss the morphology and extent of the molecular gas in our detected galaxies in Section~\ref{sec:disc}, but note here that in most cases where the CO distribution is known, it is compact.

\section{CO detections and upper limits}
\label{sec:results}

\subsection{IRAM survey observations}
\begin{figure*}
\includegraphics[width=\columnwidth,viewport=100 60 460 720,clip=true]{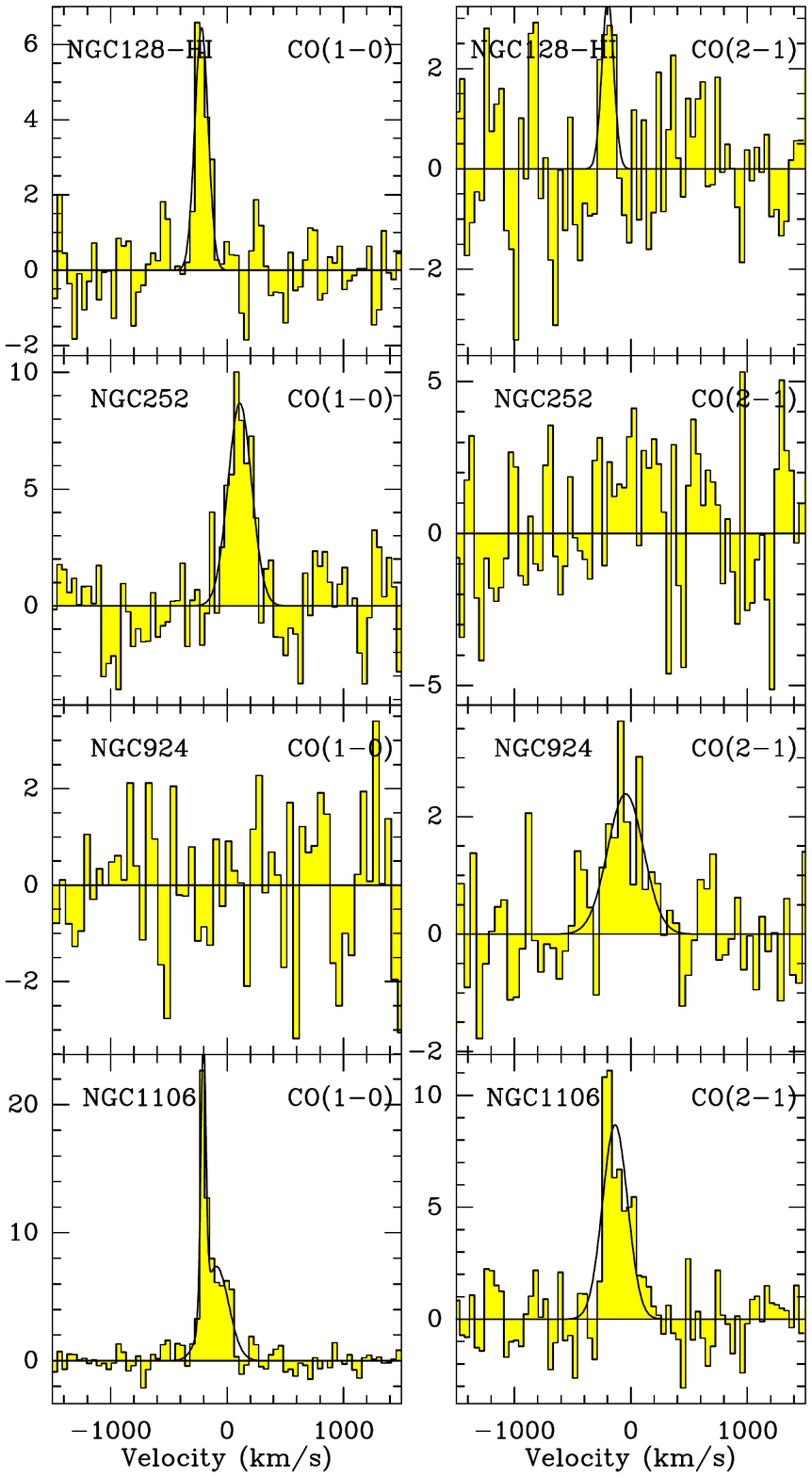}
\includegraphics[width=\columnwidth,viewport=100 60 460 720,clip=true]{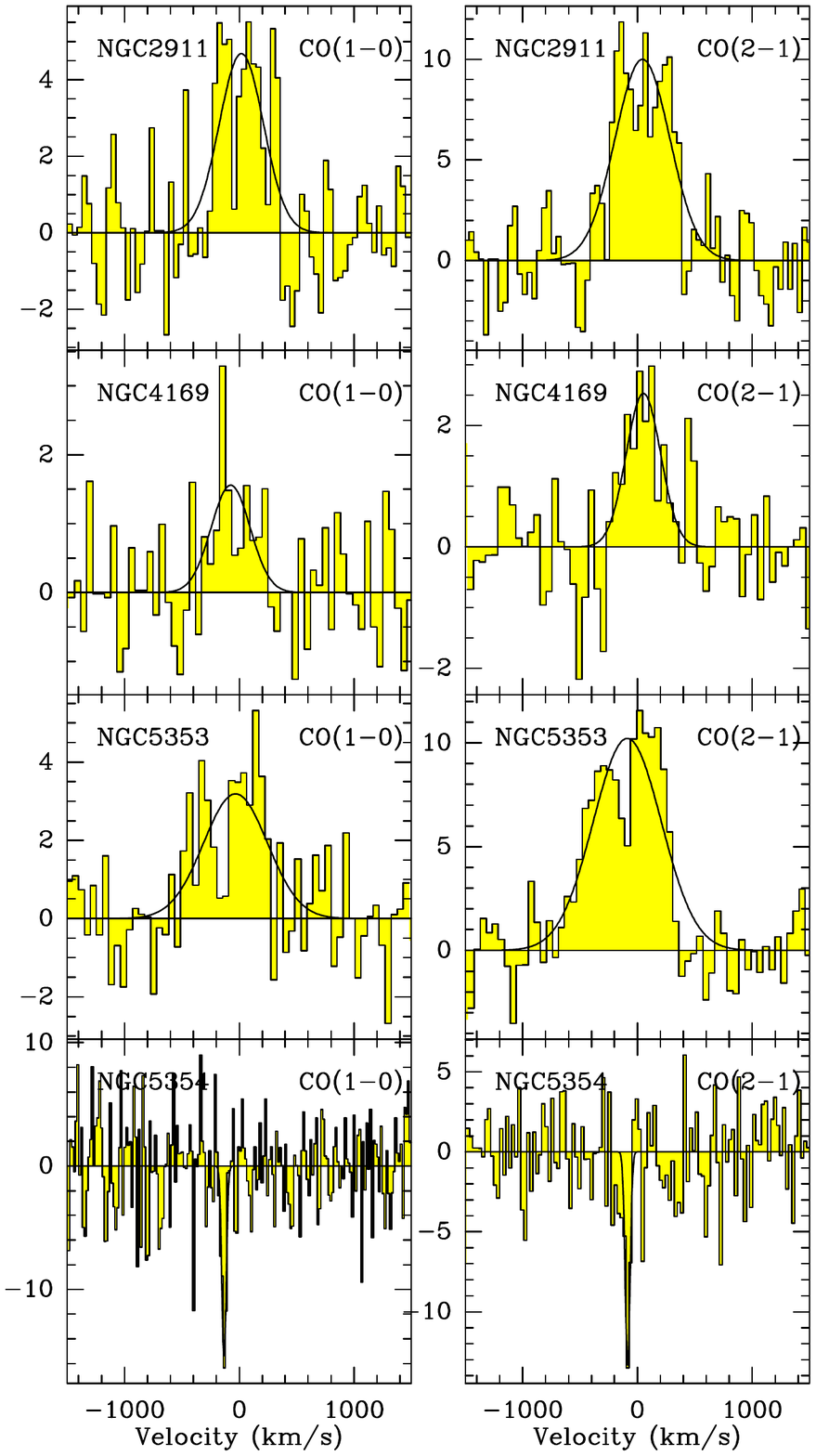}
\caption{\label{fig:IRAMdet}Spectra for the eight galaxies detected in our IRAM 30m CLoGS survey observations. For each galaxy, a pair of spectra are shown, CO(1-0) on the left, CO(2-1) on the right. The vertical axes indicate the main beam temperature \Tmb\ in mK. The best-fitting Gaussian (or in the case of NGC~1106, two-Gaussian model) is marked where lines were detected in emission or absorption. Zero velocity corresponds to the systemic velocity of each galaxy.}
\end{figure*}

Out of the 21 galaxies in our IRAM 30m survey observations of the CLoGS groups, 7 were detected at $>$4$\sigma$ significance (see Tables~\ref{tab:basic} and \ref{tab:det}). Figure~\ref{fig:IRAMdet} shows CO(1-0) and CO(2-1) spectra of each detected system. Upper limits for the undetected systems are given in Table~\ref{tab:UL}. In most cases our observations were sensitive to molecular gas mass of a few 10$^7$\Msol. 

\object{NGC~252} and \object{NGC~924} were each detected only in a single transition, CO(1-0) and CO(2-1) respectively. NGC~5353, for which we reported only an upper limit in paper I \citep[based on the results of][]{Youngetal11} was detected in both transitions in our new observations. CARMA studies have previously detected CO(1-0) emission in the galaxy with a very broad velocity dispersion \citep{Alataloetal15,Lisenfeldetal14}. In addition, we detected absorption features in both CO(1-0) and CO(2-1) close to the velocity of the neighbouring S0 \object{NGC~5354}, also shown in Figure~\ref{fig:IRAMdet} and Table~\ref{tab:det}. NGC~5354 hosts a compact ($\sim$4~mas) flat spectrum radio source with flux density $\sim$8-10~mJy between 1.4 and 15~GHz \citep{Beckeretal95,Filhoetal04,Nagaretal05}, suggesting that it is likely to be a reasonably bright source in the 100-250~GHz band. While we do not detect CO in emission in this galaxy, it is certainly plausible that clouds along the line of sight to the AGN could be detected in absorption. Assuming a conversion factor N$_{\rm H2}$/I$_{\rm CO}$=3$\times$10$^{20}$, we can estimate the column density of molecular gas along the line of sight to be N$_{\rm H2}$=2.7$\times$10$^{20}$\pcmsq.

For \object{NGC~128} we observed two locations. At the galaxy optical centroid (and optical redshift) we did not detect CO in either transition, and report the upper limits in Table~\ref{tab:UL}. However, NGC~128 is linked to its neighbor, NGC~127, by an \Hi\ bridge \citep{Chungetal12} and we therefore also observed a location between the two galaxies (RA=00 29 13.2, Dec=02 52 21), using the \Hi\ redshift drawn from the Extragalactic Distance Database all digital \Hi\ catalog \citep{Courtoisetal09}. CO was detected in both transitions at this position, and these results are shown in Table~\ref{tab:det}, in the entry NGC~128-HI. We have treated these detections as being associated with the galaxy (e.g., in Table~\ref{tab:basic}), but note the $\sim$130\kmps\ offset.

For four of the galaxies, there are suggestions of structure in the CO line profiles. NGC~5353 shows a clear double peaked line in both transitions, as expected since the gas lies in a rotating disk. \object{NGC~2911} has a similar doubly or even triply peaked profile, most clearly seen in the CO(2-1) line, and is known to host a central 500~pc scale H$_2$ disk \citep{Muller-Sanchezetal13}. \object{NGC~1106} and perhaps NGC~924 show asymmetric line profiles, indicating that the molecular gas is dynamically resolved in the IRAM~30m beam. These may indicate rotation, but could also be produced by more complex structures such as filaments or offset clouds. More detailed mapping would be needed to determine the origin of the asymmetries.

\begin{table*}[!ht]
      \caption[]{Molecular data and stellar mass of the newly detected galaxies.
For each object observed with the IRAM 30m, the first line displays the CO(1-0) and the second line the CO(2-1) results.}
\label{tab:det}
\centering
\begin{tabular}{lccccccc}
\hline\hline
Galaxy & Line & I$_{\rm CO}$ &  V & $\Delta$V$^{a}$ & T$_{mb}^{b}$ & L$^\prime_{\rm CO}$/10$^{8}$& M(H$_2$)$^{c}$\\
 & &  K km/s &   km/s &   km/s  & mK &   K.km/s.pc$^2$ &  10$^8$\Msol\\
\hline
\multicolumn{8}{l}{\textit{IRAM 30m observations}}\\
NGC~924 & CO(1-0) & $<$0.49 & - & 300 & 1.34 rms & $<$0.24 & $<$1.11 \\
          & CO(2-1) &  0.92$\pm$0.18 &  -44.3$\pm$33.4 & 363.1$\pm$86.6 &  2.4$\pm$0.8 &  0.11$\pm$0.02 &  0.52$\pm$0.10 \\ 
 NGC~4169 & CO(1-0) &  0.81$\pm$0.19 &  -75.2$\pm$51.3 & 399.6$\pm$98.8 &  1.9$\pm$0.9 &  0.31$\pm$0.07 &  1.44$\pm$0.34 \\ 
          & CO(2-1) &  0.95$\pm$0.17 &   52.9$\pm$32.6 & 353.7$\pm$75.3 &  2.5$\pm$0.5 &  0.09$\pm$0.02 &  0.42$\pm$0.08 \\ 
 NGC~5353 & CO(1-0) &  2.71$\pm$0.36 &  -32.7$\pm$44.9 & 655.4$\pm$94.6 &  3.9$\pm$1.3 &  0.41$\pm$0.06 &  1.90$\pm$0.25 \\ 
          & CO(2-1) &  7.61$\pm$0.43 &  -87.8$\pm$20.9 & 700.1$\pm$40.7 & 10.2$\pm$1.6 &  0.29$\pm$0.02 &  1.33$\pm$0.08 \\ 
 NGC~5354 & CO(1-0) & -0.90$\pm$0.18 & -132.8$\pm$ 4.6 &  45.0$\pm$11.0 & -18.9$\pm$2.9 & -0.14$\pm$0.03 & - \\ %-0.63$\pm$0.13 \\ 
          & CO(2-1) & -0.71$\pm$0.15 &  -84.7$\pm$ 5.6 &  49.7$\pm$12.7 & -13.5$\pm$0.5 & -0.03$\pm$0.01 & - \\ %-0.12$\pm$0.03 \\ 
NGC~128-HI$^d$ & CO(1-0) &  0.92$\pm$0.10 & -218.9$\pm$ 7.1 & 134.6$\pm$16.0 &  6.5$\pm$0.8 &  0.38$\pm$0.04 &  1.76$\pm$0.18 \\ 
          & CO(2-1) &  0.43$\pm$0.13 & -197.1$\pm$19.0 & 118.7$\pm$29.8 &  3.4$\pm$0.7 &  0.04$\pm$0.01 &  0.20$\pm$0.06 \\ 
  NGC~252 & CO(1-0) &  2.21$\pm$0.28 &  110.4$\pm$14.6 & 239.0$\pm$34.7 &  8.7$\pm$1.3 &  1.37$\pm$0.17 &  6.29$\pm$0.79 \\ 
 & CO(2-1) & $<$0.75 & - & 300 & 2.03 rms & $<$0.11 & $<$0.53\\
 NGC~1106 & CO(1-0) &  3.07$\pm$0.15 & -133.3$\pm$ 2.8 & 250.8$\pm$17.6 & 21.2$\pm$0.8 &  1.48$\pm$0.07 &  6.81$\pm$0.33 \\ 
          & CO(2-1) &  2.34$\pm$0.22 & -134.7$\pm$13.2 & 252.6$\pm$29.7 &  8.7$\pm$1.3 &  0.28$\pm$0.03 &  1.30$\pm$0.12 \\ 
 NGC~2911 & CO(1-0) &  2.27$\pm$0.26 &   18.5$\pm$27.6 & 455.2$\pm$50.4 &  4.7$\pm$1.3 &  0.58$\pm$0.07 &  2.66$\pm$0.31 \\ 
          & CO(2-1) &  6.00$\pm$0.42 &   44.5$\pm$20.2 & 563.0$\pm$41.1 & 10.0$\pm$1.9 &  0.38$\pm$0.03 &  1.75$\pm$0.12 \\ 
NGC~4636 & CO(1-0) & $<$0.22 & - & 300 & 0.61 rms & $<$0.005 & $<$0.023 \\
          & CO(2-1) &  0.38$\pm$0.12 &   88.6$\pm$56.7 & 304.5$\pm$98.7 &  1.2$\pm$0.6 &  0.002$\pm$0.001 &  0.010$\pm$0.003 \\ 
%          & CO(2-1) &  0.38$\pm$0.12 &   88.6$\pm$56.7 & 304.5$\pm$98.7 &  1.2$\pm$0.6 &  0.005$\pm$0.001 &  0.02$\pm$0.01 \\ 
 NGC~5846 & CO(1-0) &  0.46$\pm$0.20 &  -80.1$\pm$134.3 & 581.8$\pm$377.1 &  0.7$\pm$0.6 &  0.03$\pm$0.01 &  0.14$\pm$0.06 \\ 
          & CO(2-1) &  0.47$\pm$0.14 &  -75.9$\pm$51.3 & 334.9$\pm$91.9 &  1.3$\pm$0.7 &  0.008$\pm$0.002 &  0.04$\pm$0.01 \\ 
\multicolumn{8}{l}{\textit{APEX observations}}\\
ESO~507-25 & CO(2-1) &  1.91$\pm$0.25 &   18.0$\pm$33.0 & 472.0$\pm$65.0 &  2.4$\pm$0.9 &  0.92$\pm$0.12 &  4.23$\pm$0.56 \\ 
NGC~1779   & CO(2-1) &  2.13$\pm$0.28 &   22.0$\pm$29.0 & 434.0$\pm$55.0 &  4.7$\pm$1.1 &  0.99$\pm$0.13 &  4.57$\pm$0.60 \\ 
\object{NGC~7377}   & CO(2-1) &  2.00$\pm$0.19 &  -33.0$\pm$10.0 & 221.0$\pm$20.0 &  8.9$\pm$1.5 &  1.03$\pm$0.10 &  4.74$\pm$0.44 \\ 
\hline
% ULs calculated by hand, copied here in case you delete the table entries above
%NGC~924 & CO(1-0) & $<$0.49 & - & 300 & 1.34 rms & $<$0.24 & $<$1.11 \\
%NGC~252 & CO(2-1) & $<$0.75 & - & 300 & 2.03 rms & $<$0.11 & $<$0.53\\
%NGC~4636 & CO(1-0) & $<$0.22 & - & 300 & 0.61 rms & $<$0.01 & $<$0.05 \\
\end{tabular}
\tablefoot{Results of the Gaussian fits. $^{a}$ FWHM, $^{b}$ Peak brightness temperature, $^{c}$ obtained with the standard MW conversion ratio. In the two cases where only one transition was detected, the table shows the rms \Tmb\ for 50\kmps\ channels and 3$\sigma$ upper limits on other parameters, assuming $\Delta$V of 300\kmps. $^d$ Detected values for the NGC~128 offset pointing in the \Hi\ bridge.}
\end{table*}

\begin{table}
\caption{\label{tab:UL}Upper limits on molecular gas mass for galaxies undetected in CO(1-0) and CO(2-1).}
\centering
\begin{tabular}{lcccc}
\hline\hline\\[-3mm]
Galaxy & Line & rms & L$^\prime_{\rm CO}$/10$^8$ & M(H$_2$) \\
       &      & (mK) & (K km s$^{-1}$ pc$^{-2}$) & (10$^8$\Msol) \\
\hline\\[-1.5mm]
\multicolumn{5}{l}{\textit{IRAM 30m observations}}\\
NGC~128  & CO(1-0) &  0.9 & $<$0.13 & $<$0.60 \\ 
         & CO(2-1) &  0.9 & $<$0.04 & $<$0.16 \\ 
NGC~410  & CO(1-0) &  1.0 & $<$0.25 & $<$1.17 \\ 
         & CO(2-1) &  1.2 & $<$0.08 & $<$0.37 \\ 
NGC~584  & CO(1-0) &  1.5 & $<$0.02 & $<$0.11 \\ 
         & CO(2-1) &  1.2 & $<$0.005 & $<$0.02 \\ 
NGC~978  & CO(1-0) &  0.7 & $<$0.15 & $<$0.70 \\ 
         & CO(2-1) &  0.8 & $<$0.04 & $<$0.19 \\ 
NGC~1453 & CO(1-0) &  1.3 & $<$0.17 & $<$0.78 \\ 
         & CO(2-1) &  3.0 & $<$0.09 & $<$0.43 \\ 
NGC~1550 & CO(1-0) &  0.9 & $<$0.10 & $<$0.47 \\ 
         & CO(2-1) &  1.4 & $<$0.04 & $<$0.19 \\ 
NGC~2563 & CO(1-0) &  1.1 & $<$0.20 & $<$0.93 \\ 
         & CO(2-1) &  1.2 & $<$0.06 & $<$0.26 \\ 
NGC~3325 & CO(1-0) &  1.2 & $<$0.34 & $<$1.57 \\ 
         & CO(2-1) &  1.7 & $<$0.12 & $<$0.55 \\ 
NGC~4008 & CO(1-0) &  1.2 & $<$0.16 & $<$0.73 \\ 
         & CO(2-1) &  1.7 & $<$0.06 & $<$0.26 \\ 
NGC~4261 & CO(1-0) &  2.1 & $<$0.08 & $<$0.37 \\ 
         & CO(2-1) &  1.2 & $<$0.01 & $<$0.06 \\ 
NGC~4956 & CO(1-0) &  0.7 & $<$0.16 & $<$0.75 \\ 
         & CO(2-1) &  0.9 & $<$0.05 & $<$0.24 \\ 
NGC~5444 & CO(1-0) &  0.9 & $<$0.14 & $<$0.64 \\ 
         & CO(2-1) &  1.4 & $<$0.06 & $<$0.26 \\ 
NGC~5490 & CO(1-0) &  0.6 & $<$0.14 & $<$0.64 \\ 
         & CO(2-1) &  1.2 & $<$0.07 & $<$0.33 \\ 
NGC~5629 & CO(1-0) &  1.0 & $<$0.20 & $<$0.91 \\ 
         & CO(2-1) &  1.7 & $<$0.09 & $<$0.40 \\ 
NGC~6658 & CO(1-0) &  0.9 & $<$0.15 & $<$0.71 \\ 
         & CO(2-1) &  1.1 & $<$0.05 & $<$0.23 \\ 
NGC~5813 & CO(1-0) &  0.6 & $<$0.02 & $<$0.11 \\ 
         & CO(2-1) &  0.9 & $<$0.01 & $<$0.04 \\ 
NGC~7619 & CO(1-0) &  0.5 & $<$0.06 & $<$0.29 \\ 
         & CO(2-1) &  1.2 & $<$0.04 & $<$0.19 \\ 
%NGC~7619 & CO(1-0) &  1.2 & $<$0.16 & $<$0.73 \\ 
%         & CO(2-1) &  1.6 & $<$0.05 & $<$0.23 \\ 
\multicolumn{5}{l}{\textit{APEX observations}}\\
NGC~2292 & CO(2-1) &  1.3 & $<$0.10 & $<$0.47 \\ 
NGC~3923 & CO(2-1) &  1.6 & $<$0.06 & $<$0.29 \\ 
NGC~5061 & CO(2-1) &  1.3 & $<$0.09 & $<$0.43 \\ 
NGC~5153 & CO(2-1) &  1.2 & $<$0.38 & $<$1.75 \\ 
NGC~5903 & CO(2-1) &  1.5 & $<$0.13 & $<$0.58 \\ 
NGC~1395 & CO(2-1) &  1.3 & $<$0.06 & $<$0.27 \\ 
NGC~3078 & CO(2-1) &  0.5 & $<$0.05 & $<$0.23 \\ 
NGC~5084 & CO(2-1) &  0.7 & $<$0.03 & $<$0.16 \\ 
\hline 
\end{tabular}
\tablefoot{The rms values are in \Tmb\ in channels of 50\kmps. The upper limits on L$^\prime_{\rm CO}$ and M(H$_2$) are at 3$\sigma$ and assume $\Delta$V=300\kmps. Note that the upper limits reported for NGC~128 reflect measurements in the galaxy centre.}
\end{table}

\subsection{IRAM deep observations}
\begin{figure}
\includegraphics[width=\columnwidth,viewport=100 375 460 720,clip=true]{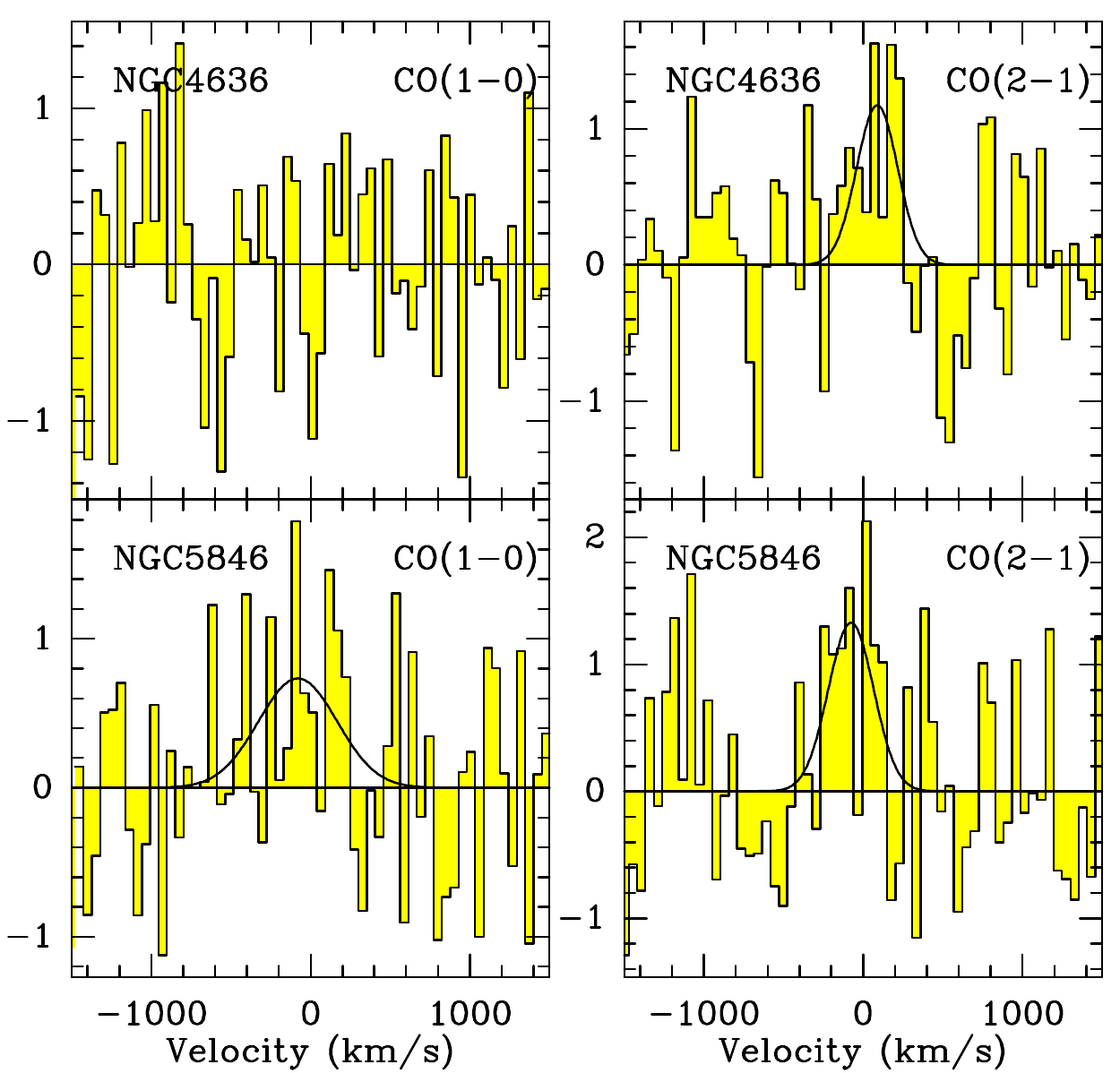}
\caption{\label{fig:IRAMdeep}Spectra for the galaxies detected in our deep IRAM 30m observations. The vertical axes indicate \Tmb\ in mK. Zero velocity corresponds to the systemic velocity of each galaxy.}
\end{figure}

Of the four galaxies for which deep IRAM 30m observations were performed, two were detected, NGC~4636 and NGC~5846. Our previous short observation of NGC~5846 had provided only an upper limit (see paper I). Figure~\ref{fig:IRAMdeep} shows spectra for the two galaxies, with NGC~5846 detected in both transitions and NGC~4636 only in CO(2-1). The gas masses detected in these galaxies (Table~\ref{tab:det}) are a factor of $\sim$10-100 less than those typically detected in our survey observations, and in the undetected systems we place upper limits on the molecular gas content of $\sim$10$^7$\Msol\ (Table~\ref{tab:UL}). For NGC~7619, which was undetected in both survey and deep observations, the upper limits reported in Tables~\ref{tab:basic} and \ref{tab:UL} were determined from the combination of the two datasets.

\citet{Temietal17} report CO(2-1) ALMA 12m array observations of NGC~4636 and NGC~5846, with measured molecular gas masses of 2$\times$10$^5$\Msol\ and 2.1$\times$10$^6$\Msol\ respectively. Comparing CO(2-1) line intensities (in units of Jy~km~s$^{-1}$, to avoid any differences in assumed distance or mass conversion factor) we find that the IRAM 30m detects a factor of $\sim$7 more CO(2-1) emission in NGC~4636 and a factor of $\sim$2 more in NGC~5846. This indicates that a significant fraction of the molecular gas is resolved out in the ALMA observations. ALMA 7m array observations (David et al., in prep.) confirm that this is also the case in NGC~5044, where the total CO flux observed in the IRAM~30m beam is recovered when larger angular scales are included.

\subsection{APEX observations}
\begin{figure*}
\centerline{
\includegraphics[height=0.8\textwidth,angle=-90,viewport=340 30 540 570,clip=true]{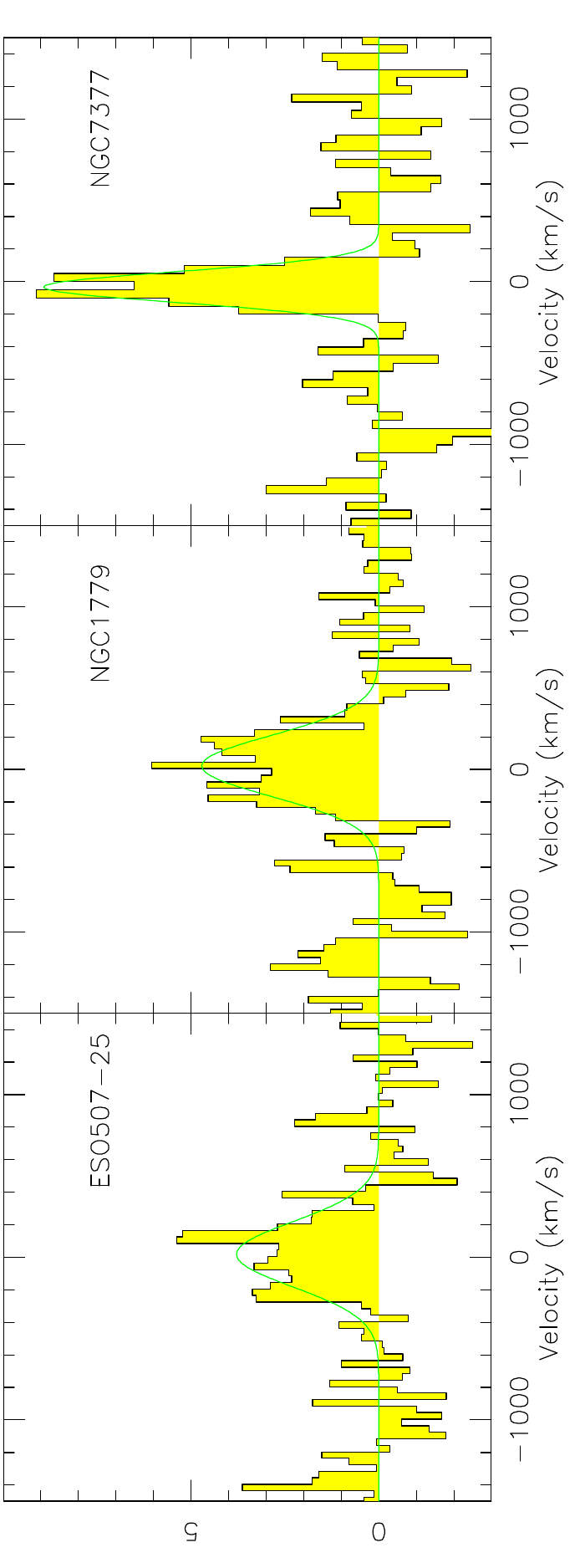}
}
\caption{\label{fig:APEX}CO(2-1) spectra for the galaxies detected in our APEX observations. The vertical axes indicate \Tmb\ in mK. Zero velocity corresponds to the systemic velocity of each galaxy.}
\end{figure*}

Of the 11 APEX targets, 3 were detected. Figure~\ref{fig:APEX} shows the CO(2-1) spectra for these galaxies, and their fluxes, estimated molecular gas masses and other properties are listed in Tables~\ref{tab:basic} and \ref{tab:det}. These systems contain 3-4$\times$10$^8$\Msol\ of molecular gas, at the upper end of the range of gas masses seen in the sample. Upper limits on the undetected systems are shown in Table~\ref{tab:UL}, and show that our APEX observations were sensitive to gas masses of a few 10$^7$\Msol, comparable to the IRAM 30m observations.

\subsection{CLoGS detection fraction}
In total, 21 of the 53 CLoGS BGEs were found to contain molecular gas, giving a detection rate 40$\pm$9\%. This is in good agreement with the detection fraction found in paper~I (43$\pm$14\%). The detection rate for early-type galaxies found from the ATLAS$^\mathrm{3D}$ sample is 22$\pm$3\% \citep{Youngetal11}, with CO sensitivities similar to our survey observations (see paper~I for a comparison). Although the uncertainties mean that the difference is only significant at the $\sim$2$\sigma$ level, this hints that group-dominant early-type galaxies may be more likely to contain molecular gas than the general population. It is notable that this is the opposite of the general trend for early-type galaxies in denser environments to be cold-gas-poor compared to those in the field. 

\subsection{CO line ratios}
For those galaxies observed with the IRAM~30m and detected in both transitions we can calculate the CO(2-1)/CO(1-0) line ratio in brightness temperature, which is related to both the excitation and distribution of the gas. We find values in the range $\sim$0.5-3, with the majority of values $>$1. Since the CO(2-1) beam size is a factor 4 smaller than that at CO(1-0), we would expect line ratios of 4 if the gas is optically thick and compact compared to the smaller beam (i.e., it approximates a point source), and 1 if it is evenly distributed over an area greater than both beams. However, we might expect ratios $<$1 if the gas is subthermally excited \citep{BraineCombes92}. Given the range of line ratios we observe, we can only say that the gas distribution is likely to vary significantly among our galaxies.

\subsection{Overview: CO}
In total, we report CO detections for 12 BGEs, and upper limits for a further 24, from observations with a typical sensitivity of a few 10$^7$\Msol. For the 53-group CLoGS sample, we find that 40$\pm$9\% of BGEs are detected in CO. This is marginally higher than the detection rate for the general population of early-type galaxies.

\section{Analysis}
\label{sec:analysis}

In this section we consider results for our BGE sample as a whole, including those galaxies described in paper~I.

\subsection{H\textsc{i} and molecular gas}
\label{sec:HI}
While CLoGS has not been surveyed in \Hi, measurements of the atomic gas mass in and around many of the galaxies are available from the literature. Table~\ref{tab:basic} lists \Hi\ masses and limits for our galaxies, and we have also updated the \Hi\ masses listed in paper~I (see Table~\ref{tab:bigtab}). In total, 27 of the 53 CLoGS BGEs are detected in \Hi\ (as is NGC~4636) and upper limits are available for a further 14 (plus NGC~5813). The measurements are not in any sense statistically complete, having been made with a variety of instruments, with different sensitivities and spatial resolutions, and covering only a subset of our BGEs. Whereas the CO observations generally only cover the galaxy cores, the \Hi\ measurements usually include the whole galaxy, and may capture material in the outskirts, whose molecular content (if any) would not have been included in the CO measurements. We therefore treat the \Hi\ masses with caution, but consider them of interest for the insight they may give into the size of the cold gas reservoirs in our galaxies. The literature sources provide uncertainties on the H\textsc{i} mass for only $\sim$40\% of the galaxies and we therefore do not include them, but note that in most cases where an uncertainty was given, it was $\le$0.1dex.

For four systems, we find conflicting H\textsc{i} measurements in the literature. For NGC~940, the Nan\c{c}ay radio telescope finds a strong, double-peaked detection \citep{Patureletal03b}, but the galaxy falls in the footprint of the Arecibo Legacy Fast ALFA Survey \citep[ALFALFA,][]{Haynesetal18} and, based on the flux in the earlier measurement, should be detectable. The Nan\c{c}ay beam likely includes the spiral galaxy UGC~1963 which has a similar velocity and may contribute to the measured flux, but we note that UGC~1963 is separately detected with a narrower linewidth. NGC~4636 and NGC~5846 both have H\textsc{i} detections from single-dish telescopes \citep{BottinelliGouguenheim77,BottinelliGouguenheim79,Knappetal78} but also non-detections \citep{HuchtmeierRichter89,DuprieSchneider96}. Both would be expected to be detected by ALFALFA but are not, and \citet{Youngetal18} were unable to detect either galaxy with the VLA, though very diffuse H\textsc{i} could perhaps have been resolved out. NGC~7619 is similar, with both detections \citep{Serraetal08} and non-detections \citep{Huchtmeier94} reported. The galaxy is not reported in the ALFALFA catalog, but its reported flux (0.04~Jy~km~s$^{-1}$) falls well below the survey detection limit \citep{Haynesetal11}. We have flagged the H\textsc{i} masses for these four galaxies as meriting particular caution in Table~\ref{tab:basic} (and Appendix~\ref{bigtab}).

Our finding that 51$\pm$10\% of CLoGS BGEs contain \Hi\ is comparable to the ATLAS$^{3D}$ result for early type galaxies outside the Virgo cluster, 39$\pm$6\% \citep{Serraetal12}. However, our detection fraction is greater than that for their Virgo cluster members, 10$\pm$5\%, at 3.7$\sigma$ significance. Given the incompleteness of our \Hi\ data, our detection fraction can be considered a lower limit, so this difference is not a product of the uneven \Hi\ coverage.

\begin{figure}
\includegraphics[width=\columnwidth,viewport=20 200 575 760,clip=true]{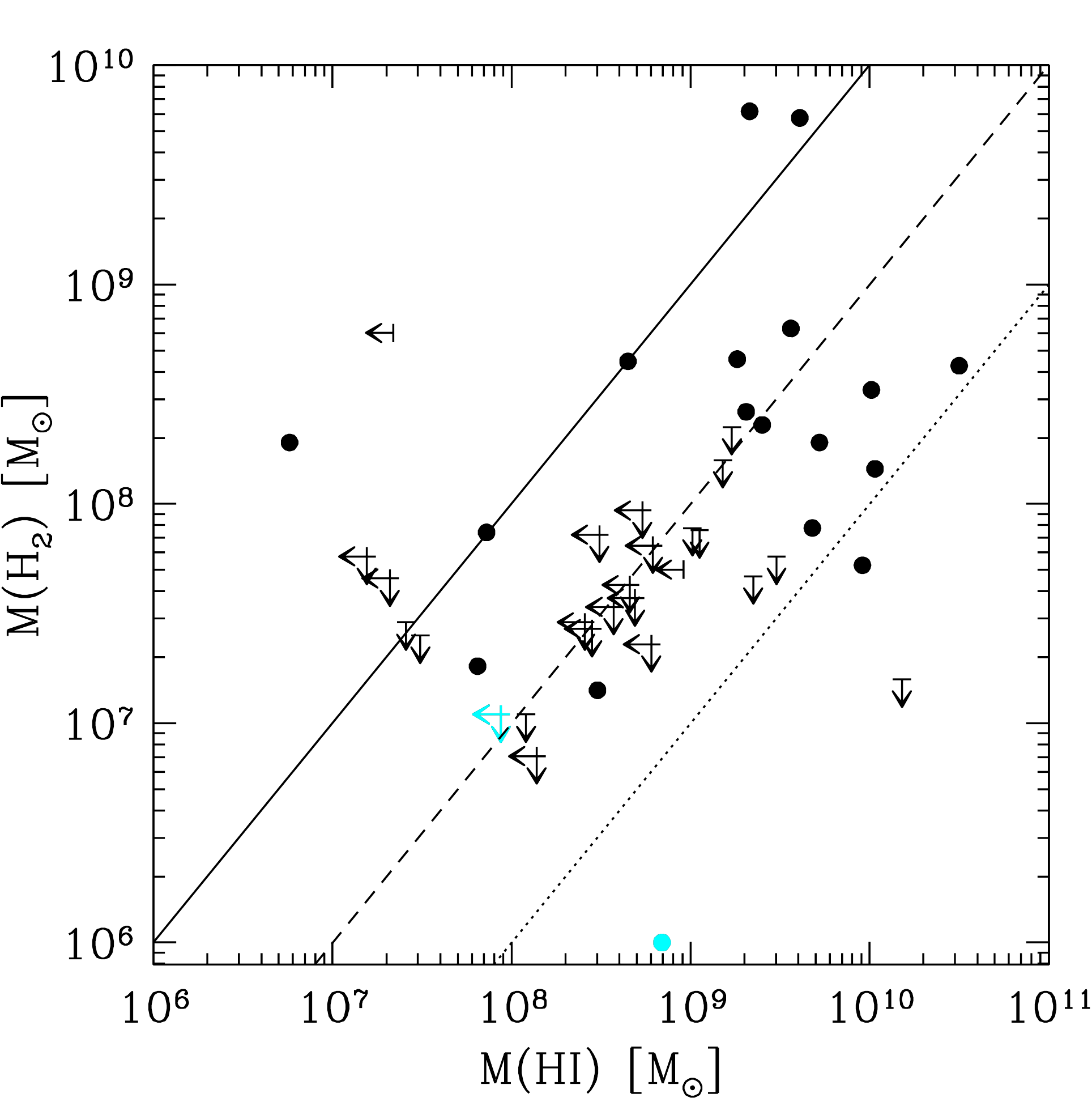}
\caption{\label{fig:HIH2} Molecular vs atomic gas masses in our galaxies. Black points indicate CLoGS BGEs, cyan points non-CLoGS galaxies. The solid, dashed, and dotted lines indicate M(H$_2$)/M(\Hi) ratios of 1, 0.1 and 0.01 respectively.}
\end{figure}

\subsubsection{Molecular to atomic gas mass ratios}
Figure~\ref{fig:HIH2} shows the molecular and atomic gas masses for our galaxies plotted against each other. The majority of the galaxies have molecular to atomic gas ratios M(H$_2$)/M(\Hi)=0.01-1. Galaxies falling at the low end of this range are the most likely to have values biased by the difference in coverage between \Hi\ and CO. NGC~4636 is the most extreme case, with only 10$^6$\Msol\ of molecular gas, but $\sim$7$\times$10$^8$\Msol\ of \Hi. The \Hi\ profile has a broad line width (570$\pm$100\kmps) and has two peaks, perhaps suggesting rotation. \Hi\ mapping would be needed to determine whether the \Hi\ is cospatial with the molecular gas, or (as a broader linewidth would suggest) is at larger radii, for example in a disk or ring which does not extend in to the galaxy core. However, we note again that the H\textsc{i} measurements in the literature are conflicting, and that VLA mapping did not detect any \Hi.

Galaxies with high M(H$_2$)/M(\Hi) are less likely to be affected by the different beam sizes. NGC~524 has the highest ratio among the systems where both gas phases are detected, M(H$_2$)/M(\Hi)=33. Its H$_2$ and \Hi\ are definitely not cospatial; \citet{Oosterlooetal10} find only a small \Hi\ cloud offset $\sim$1\arcm\ to the north side of the galaxy core, outside the beam of our CO observation. NGC~3665 has M(H$_2$)/M(\Hi)$>$27.5, and has already been shown to contain a compact molecular gas disk \citep{Alataloetal13} with an exceptionally high gas density compared to the star formation rate density \citep{Davisetal14}. Herschel [C\textsc{ii}] observations confirm the presence of large quantities of molecular gas, and the suppressed star formation rate \citep{Xiaoetal18}. NGC~940 and NGC~7252 also have M(H$_2$)/M(\Hi)$>$1, indicating a large fraction of dense gas in their disks.

\subsubsection{Cold gas masses}
We compute the total cold gas mass M(\Hi+H$_2$) for the 31 galaxies (30 CLoGS BGEs) with detected molecular and/or atomic gas. Examining the ratio of cold gas mass to stellar mass, M(\Hi+H$_2$)/M$_*$, we find values in the range 0.001-0.360, with a mean value of $\sim$5 per cent, though this is likely to be biased high since we only have upper limits for a number of our systems. The mean is dominated by the \Hi\ masses; the molecular gas to stellar mass ratio has a range 3$\times$10$^{-4}$-7$\times$10$^{-2}$ and a mean value $\sim$1 per cent. In general, the range of \Hi\ fractions is comparable to that found for the ATLAS$^{3D}$ sample \citep{Serraetal12}.

\subsubsection{Overview: H\textsc{i}}
Drawing on data from the literature, we find that at least 49$\pm$10\% of the CLoGS BGEs contain \Hi, comparable to the fraction in the general population of early type galaxies but significantly greater than the fraction for Virgo cluster early-types. Combining \Hi\ and CO, we find that 57$\pm$10\% of CLoGS BGEs contain cold gas, with a wide range of M(H$_2$)/M(\Hi) ratios.

\subsection{Star formation and nuclear activity}
\label{sec:SFR}

As described in paper~I, we calculated far infrared (FIR) luminosities and dust masses based on IRAS or \textit{Spitzer} fluxes, for the galaxies where these were available. The resulting values are listed in Table~\ref{tab:basic}.

%As in paper~I, we calculated far infrared (FIR) luminosities from IRAS 60 and 100~$\mu$m fluxes where possible, or from \textit{Spitzer}/MIPS 24, 70 and 160~$\mu$m fluxes if IRAS fluxes were unavailable. The FIR flux ratios were also used to estimate the dust mass

%\begin{equation}
%M_{\rm d} = 4.8\times10^{-11}\frac{S_{\nu o}\rm{D}_{\rm Mpc}^2}{(1+z)\kappa_{\nu \rm{r}}B_{\nu \rm{r}}(T_{\rm d})}M_\odot,
%\end{equation}

%where $S_{\nu o}$ is the observed FIR flux in Jy, D$_{\rm Mpc}^2$ is the luminosity distance in Mpc, $B_{\nu \rm{r}}$ the Planck function at the rest frequency $\nu{\rm r}$=$\nu$o(1=$z$), and $\kappa_{\nu{\rm r}}$ is the rest-frame mass opacity of the dust, for which we adopt a value of 25~cm$^2$~g$^{-1}$ \citep{Hildebrand83,Dunneetal00,Draine03}.

For galaxies with measured FIR luminosities, we calculate the star formation rate (SFR) from the relation of \citet{Kennicutt98}, SFR=L$_{\rm FIR}$/(5.8$\times$10$^9$\Msol). We then estimate the gas depletion timescale, $\tau$=5.8~M(H$_2$)/L$_{\rm FIR}$ Gyr, where the molecular gas mass and FIR luminosity are in solar units. Table~\ref{tab:SFR} lists star formation rates and depletion timescales for our targets, as well as the specific star formation rate (sSFR), defined as SFR/M$_*$, where M$_*$ is the stellar mass of the galaxy. We determine M$_*$ using a mass-to-light ratio from the models of \citet{BellDeJong01}, based on galaxy colours from Sloan Digital Sky Survey $ugriz$ and 2~Micron All Sky Survey $JHK$ magnitudes.

\begin{table}
\caption{\label{tab:SFR}Star formation rates and depletion times}
\centering
\begin{tabular}{lccc}
\hline\hline
Galaxy & SFR & $\tau$ & sSFR \\
       & (\Msolpyr) & (Gyr) & (Gyr$^{-1}$) \\
\hline
    NGC~128 & 1.193$\pm$0.085 & 0.15$\pm$0.02 & 9.7e-03 \\ 
    NGC~252 & 1.193$\pm$0.146 & 0.53$\pm$0.07 & 1.0e-02 \\ 
    NGC~584 & 0.024$\pm$0.004 & -            & 5.8e-04 \\ 
    NGC~924 & 0.545$\pm$0.381 & 0.10$\pm$0.02 & 9.5e-03 \\ 
    NGC~978 & 0.157$\pm$0.032 & -            & 1.3e-03 \\ 
   NGC~1106 & 1.847$\pm$0.274 & 0.37$\pm$0.02 & 2.5e-02 \\ 
   NGC~1395 & 0.023$\pm$0.005 & -            & 2.8e-04 \\ 
   NGC~1453 & 0.217$\pm$0.032 & -            & 1.5e-03 \\ 
   NGC~1779 & 0.926$\pm$0.187 & 0.49$\pm$0.06 & 1.6e-02 \\ 
   NGC~2292 & 0.314$\pm$0.022 & -            & 5.1e-03 \\ 
   NGC~2911 & 0.117$\pm$0.020 & 2.28$\pm$0.27 & 1.8e-03 \\ 
   NGC~3325 & [2.17$\pm$0.09]$\times$10$^{-6}$ & -            & 2.9e-08 \\ 
   NGC~3923 & 0.004$\pm$0.001 & -            & 3.9e-05 \\ 
   NGC~4008 & 0.128$\pm$0.075 & -            & 1.6e-03 \\ 
   NGC~4169 & 4.972$\pm$0.116 & 0.03$\pm$0.01 & 6.4e-02 \\ 
   NGC~4261 & 0.034$\pm$0.010 & -            & 3.1e-04 \\ 
   NGC~4636 & 0.015$\pm$0.001 & 0.07$\pm$0.02 & 4.6e-04 \\ 
   NGC~4956 & 0.598$\pm$0.138 & -            & 7.4e-03 \\ 
   NGC~5061 & 0.004$\pm$0.001 & -            & 4.7e-05 \\ 
   NGC~5084 & 0.128$\pm$0.016 & -            & 1.6e-03 \\ 
   NGC~5353 & 0.095$\pm$0.009 & 2.01$\pm$0.26 & 8.4e-04 \\ 
   NGC~5813 & 0.021$\pm$0.003 & -            & 2.7e-04 \\ 
   NGC~5846 & 0.019$\pm$0.003 & 0.72$\pm$0.31 & 2.9e-04 \\ 
   NGC~7377 & 0.509$\pm$0.062 & 0.93$\pm$0.09 & 4.6e-03 \\ 
   NGC~7619 & 0.255$\pm$0.081 & -            & 1.6e-03 \\ 
  ESO507-25 & 0.497$\pm$0.061 & 0.85$\pm$0.11 & 5.6e-03 \\ 

\hline
\end{tabular}
\end{table}

The gas depletion timescales only account for the detected molecular gas, and as in paper~I, we find typical values $<$1~Gyr, despite very low specific star formation rates. This confirms our previous conclusion that the molecular gas content of these galaxies must be replenished on relatively short timescales, whatever its origin.

\begin{figure}
\includegraphics[width=\columnwidth,viewport=20 200 575 760,clip=true]{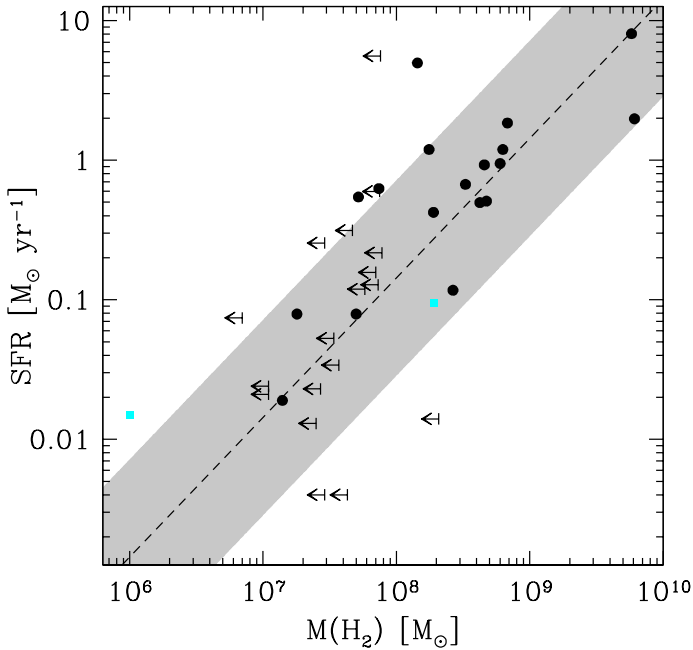}
\caption{\label{fig:SFR_MH2} FIR SFR against M(H$_2$)  for our CLoGS dominant galaxies (black circles or arrows) and non-CLoGS galaxies observed in our deep observations (cyan squares or arrows). The dashed line indicates the molecular gas mass to SFR relation of \citet{GaoSolomon04}, equivalent to typical depletion timescales $\sim$1~Gyr, and the shaded region a range of a factor of 5 about the relation.}
\end{figure}

Figure~\ref{fig:SFR_MH2} shows SFR and M(H$_2$) for the CLoGS BGEs and the non-CLoGS group-dominant ellipticals NGC~4636 and NGC~5813, compared to the SFR to molecular gas mass relation of \citet{GaoSolomon04} for normal spirals with SFR$<$20\Msolpyr, SFR=1.43$\times$10$^{-9}$M(H$_2$) \Msolpyr, where M(H$_2$) is in units of \Msol. This relation is equivalent to a depletion timescale of $\sim$0.7~Gyr, and the majority of our galaxies fall within a factor of 5 of the relation. We note that the FIR luminosities do not appear to be correlated with radio luminosities (Pearson and Spearman's rank correlations find only a $\sim$1$\sigma$ significant correlation), so we have no reason to believe that AGN are significantly impacting the SFR estimates.  The galaxies furthest from the relation include NGC~315, which hosts the most luminous radio source in our sample, B2~0055+30, but also \object{NGC~4169}, NGC~128, NGC~924, NGC~2768 and NGC~4636, all of which host only weak radio sources. Emission from dust associated with the filamentary nebulae in cooling flows \citep[as in, e.g., NGC~1275,][]{Mittaletal12} could also affect the FIR flux in some systems, but the best example of such a nebula in our sample, NGC~5044, falls very close to the relation, suggesting that the impact may be small.

%\begin{figure}
%\includegraphics[width=\columnwidth,viewport=20 200 575 760,clip=true]{LFIR_LCO.pdf}
%\caption{\label{fig:LFIR_LCO} FIR vs CO luminosities for our CLoGS dominant galaxies (black circles) and non-CLoGS galaxies observed in our deep observations (cyan squares). The shaded region and dotted line indicate the typical range of depletion timescales of nearby spiral galaxies, 2~Gyr with a factor of 3 scatter \citep{Bigieletal08}.}
%\end{figure}

If we compare the SFR to the total cold gas mass available in each galaxy, the degree of scatter increases, with depletion times ranging over 0.1-100~Gyr. In some cases, the \Hi\ is located at large radii \citep[e.g., the ring surrounding NGC~2292/2293][]{Barnes99} and is unlikely to be connected to current star formation. However, some of the longest depletion times are found in galaxies with large \Hi\ disks, such as NGC~5084 \citep{Pisanoetal11} or NGC~1167 \citep{Struveetal10}, suggesting that star formation is suppressed in these systems.

\begin{figure}
\includegraphics[width=\columnwidth,viewport=20 200 575 760,clip=true]{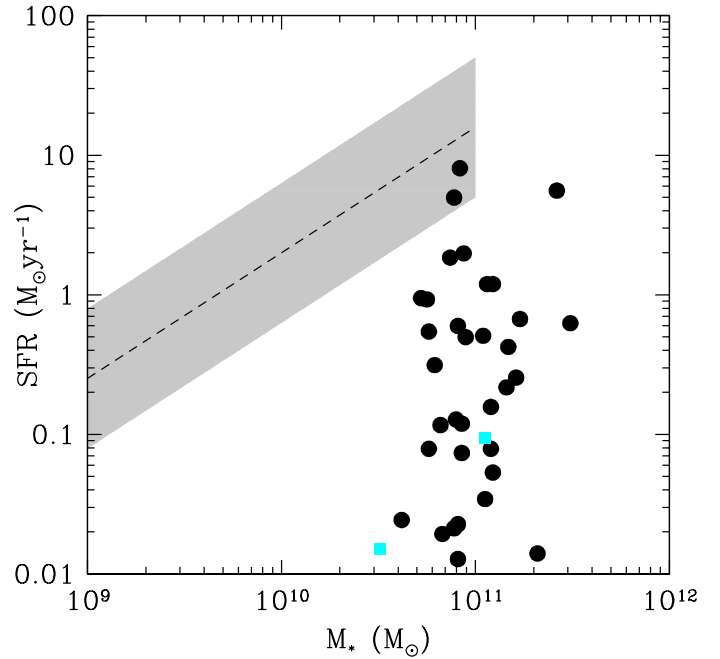}
\caption{\label{fig:SFR_Mstar} The SFR-stellar mass diagram, with CLoGS dominant galaxies marked by black circles, and non-CLoGS galaxies by cyan squares. The shaded region indicates the main sequence of star formation galaxies at $z$=0 \citep{Wuytsetal11}}
\end{figure}

Figure~\ref{fig:SFR_Mstar} shows the relation between stellar mass and SFR for galaxies detected in the FIR. Our galaxies fall in a narrow mass range ($\sim$4-30$\times$10$^{10}$\Msol) and almost all fall below the main sequence of star-forming galaxies at $z$=0 \citep{Wuytsetal11}, indicating that they are quenched, red sequence systems. A few have SFR$>$1~\Msolpyr, and the highest have star formation rates comparable to those of massive spiral galaxies. These include NGC~7252 a post-merger starburst galaxy in which high SFR is expected \citep[e.g.,][]{Schweizer82,Hibbardetal94,Dopitaetal02} and NGC~4169 in the compact group HCG~61 \citep{Hicksonetal89} where close tidal encounters may have had some impact, but also the relatively unremarkable NGC~1060.

\subsubsection{Gas mass vs radio luminosity}
A high fraction of massive early-type galaxies host radio sources and the CLoGS BGEs are no exception \citep[and references therein]{Kolokythasetal18}. Radio fluxes are available for the majority of our BGEs, from a variety of sources. These include our own 610 and 235~MHz GMRT observations for the high-richness half of the sample \citep{Kolokythasetal18}, the 1.4~GHz FIRST \citep{Beckeretal95} and NVSS \citep{Condonetal98,Condonetal02} surveys. Our 610~MHz observations (half-power beam width, HPBW=$\sim$6\arcs) and FIRST (HPBW=5.4\arcs) provide the spatial resolution necessary to confirm that sources are associated with the target galaxy, but fluxes from NVSS (HPBW=45\arcs) are preferred where available, since it is less prone to break up extended sources into separate components. 

We have compiled 1.4~GHz fluxes for our sample, and the fluxes for our new observations are listed in Table~\ref{tab:basic}. Where neither NVSS or FIRST fluxes were available, we have drawn on the VLA sample of \citet{Brownetal11}, GMRT 1.4~GHz measurements \citep{OSullivanetal18}, or (for NGC~924, NGC~980 and NGC~2563) fluxes extrapolated from our 610~MHz measurements. For NGC~5629, only a 2.38~GHz Arecibo flux is available, and we therefore estimated a 1.4~GHz flux from this value assuming a spectral index of 0.8. In total, we find that 46/53 (87$\pm$13\%) CLoGS BGEs are detected at some radio frequency, in the currently available radio data.

\begin{figure}
\includegraphics[width=\columnwidth,viewport=20 200 575 760,clip=true]{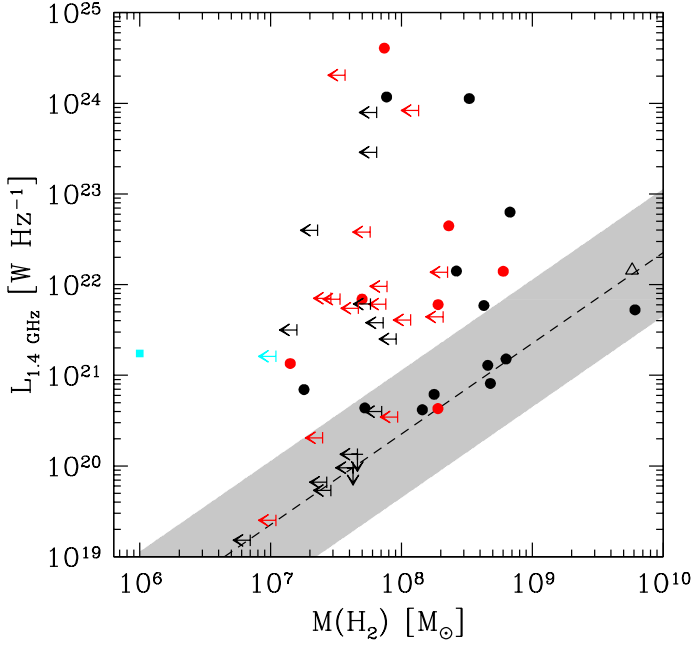}
\caption{\label{fig:Lradio} Molecular gas mass against 1.4~GHz radio continuum power for our group dominant galaxies. Points (circle, square or triangle) indicate systems detected in both radio and CO, arrows indicate 3$\sigma$ upper limits on CO mass or 2$\sigma$ limits on 1.4~GHz luminosity. The open triangle indicates the post-merger starburst galaxy NGC~7252. Red points indicate CLoGS galaxies whose groups have measured X-ray luminosities $>$10$^{41}$\ergps, black points X-ray faint or unmeasured CLoGS systems, and cyan points non-CLoGS galaxies, both of which are X-ray bright. The black line represents the expected radio emission from star formation, assuming the molecular gas mass to star formation rate relation of \protect\citet{GaoSolomon04}, with the grey shaded region indicating a range of a factor$\pm$5.}
\end{figure}

Figure~\ref{fig:Lradio} shows 1.4~GHz radio luminosity plotted against molecular gas mass for our sample. Following \citet{Nylandetal17}, we estimate the expected radio emission from star formation based on the molecular gas mass to star formation rate relation of \citet{GaoSolomon04} and the L$_{\rm 1.4GHz}$:SFR relation of \citet{Murphyetal11}. This is shown as a dashed line in Figure~\ref{fig:Lradio}, with a factor of $\pm$5 uncertainty indicated by the shaded region. This relation appears to provide a lower bound to radio luminosity in our sample. A number of our systems fall within the band, including our two CO-richest galaxies, NGC~940 and NGC~7252. The excellent match between the expected and observed radio emission in the starburst NGC~7252 suggests that the adopted M(H$_2$):SFR relation holds for our galaxies.

Galaxies falling above the relation likely have a contribution to their radio luminosity from AGN. In many cases, the radio morphology confirms this. Kolokythas et al. describe a scheme by which they classify our group-central radio sources according to their morphology, with the following classes: large-scale jets, which extend $>$20~kpc and often beyond the host galaxy and into the surrounding IGM; small-scale jets with extent $<$20~kpc; diffuse sources, which have extended amorphous radio emission with no clear jets or lobes but which is not related to star formation; and point-like sources, which are unresolved. Using this scheme, we find that all 6 of our galaxies with L$_{\rm 1.4GHz}$$>$5$\times$10$^{23}$~W~Hz$^{-1}$ host large-scale jet sources. The 8 galaxies with L$_{\rm 1.4GHz}$=0.1-5$\times$10$^{23}$~W~Hz$^{-1}$ have a mix of radio morphologies, but 75\% are extended (excluding NGC~7252), hosting either jets or diffuse emission. This confirms that our radio-bright BGEs tend to be AGN rather than star-formation dominated.

Conversely, the 8 CO-detected galaxies whose radio luminosity is consistent with star formation all host radio point sources (again excluding NGC~7252). Based on the distance of the galaxies and the resolution of the available data, if this emission arises from star formation, the star forming regions would be limited to $\sim$1.5-10~kpc in diameter. Given the small star formation rates, this does not seem implausible.

Nyland et al. examined this relation for early-type galaxies drawn from the ATLAS$^{\rm 3D}$ survey covering a wide range of environments. They found that 5/56 ($\sim$9$\pm$4\%) CO-detected galaxies fell more than a factor of 5 above the relation, and that the fraction did not change if CO-undetected galaxies were included (24/260). A much larger fraction of our group dominant galaxies fall above the relation: 13/22 CO-detected galaxies (60$\pm$16\%) and 32/50 (64$\pm$11\%) if non-detections are included. This is a 3$\sigma$ or 5$\sigma$ significant difference respectively, strongly suggesting that our BGEs are more likely than the general population of early-type galaxies to host radio luminous AGN. While this is expected (see Kolokythas et al. for a discussion), it is notable that the difference is seen in both CO-detected and undetected galaxies.

There is no clear difference in the rate of CO detection with radio luminosity. Excluding NGC~7252 as a source we know to be SF dominated and dividing our galaxies at L$_{\rm 1.4~GHz}$=10$^{22}$~W~HZ$^{-1}$, we find that 7/14 of the more radio luminous systems are detected in CO (50$\pm$14\%) compared to 14/32 radio fainter systems (44$\pm$12\%). This suggests that the presence of molecular gas does not automatically lead to radio-luminous nuclear activity \citep[see also][]{Baldietal15}. We also find no correlation between radio power and CO(2-1)/CO(1-0) line ratio, suggesting that AGN have only a limited effect on the excitation of the molecular gas in our BGEs.

\subsubsection{Overview: star formation and AGN}
We confirm our finding from paper~I that most BGEs have low star formation rates and short depletion times. While a significantly larger fraction of BGEs are AGN-dominated than is the case for the general population of early type galaxies, molecular gas mass is not directly linked to nuclear activity.

\subsection{Group properties}
\label{sec:grp}

\subsubsection{Richness}
\begin{figure}
\includegraphics[width=\columnwidth,viewport=20 200 575 760,clip=true]{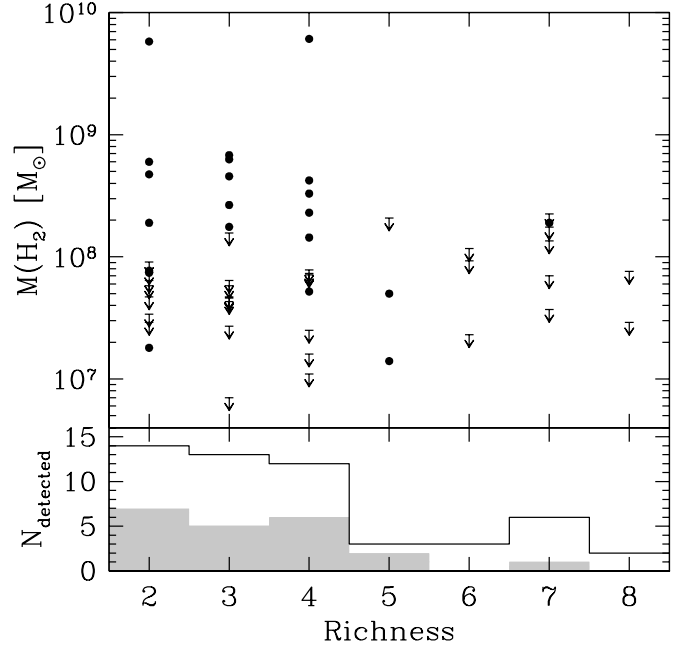}
\caption{\label{fig:R}Molecular gas mass against group richness parameter $R$, essentially the number of large galaxies in the group. The upper panel shows the individual detections and limits, while the lower panel shows the number of galaxies in each bin (black) and the number detected (grey). CO is detected in $\sim$50\% of BGEs with $R$$\leq$5.}
\end{figure}

Figure~\ref{fig:R} shows molecular gas mass of the CLoGS BGEs (i.e., excluding NGC~4636 and NGC~5813) ranked by the richness of the group, where richness $R$ is defined as the number of group members with $B$-band luminosity log L$_{\rm b}$$>$10.2 \citep{OSullivanetal17}. While we have a limited number of systems in the highest richness classes, there is some suggestion of a difference between high- and low-richness groups. For $R$=2-5, 20/42 (47.5$\pm$10.6\%) groups have a BGE with detected CO, whereas for $R$=6-8, only 1/11 (9$\pm$9\%) has a CO-detected BGE. This difference is only 2.75$\sigma$ significant, but suggestive. Richness gives an indication of the complexity of the group environment, and may be a tracer of mass. A similar trend is seen in \Hi; atomic gas is detected in 23/42 BGEs in groups with $R$=2-5, compared to 3/11 with $R$=6-8. However, while this appears supportive of the CO result, we must bear in mind that we do not have \Hi\ data for every galaxy, so the relative fractions could change.

\subsubsection{IGM X-ray luminosity}
At present, X-ray coverage of the CLoGS is incomplete. The richer half ($R$$\ge$4) of the sample has been fully characterized, but many of the poorer groups lack good-quality X-ray data. For the high-richness groups we know that there is only a weak correlation between X-ray luminosity and richness among those groups with an X-ray detected IGM, but that groups with no detected IGM tend to fall in the lowest richness ($R$=4) bin \citep{OSullivanetal17}. There is no obvious correlation between CO detection and the presence or absence of a hot IGM. Four of the nine CO-detected BGEs lie at the centres of groups with a hot IGM, five in groups with no detected IGM. However, with only nine CO-detected BGEs in the richer half of the CLoGS, we cannot draw strong conclusions. We will need to revisit the question once X-ray observations of the full sample are complete.

In Figure~\ref{fig:Lradio} we have marked the BGEs of X-ray bright groups, defined as having X-ray luminosities within R$_{500}$ $>$10$^{41}$\ergps, drawing these luminosities from our own studies \citep{OSullivanetal17,OSullivanetal18} and the literature \citep{Sunetal03,OsmondPonman04,Cavagnoloetal09}. There is no strong segregation between X-ray bright and faint groups, though it is notable that (as yet) only one BGE of an X-ray bright system contains more than 3$\times$10$^8$\Msol\ of molecular gas (NGC~3665). X-ray bright systems appear to host BGEs with a wide range of molecular gas masses and radio luminosities, again suggesting that the presence of an IGM is not a strong constraint on either property. This would be expected if interactions and mergers are an important source of cold gas for BGEs.

\begin{figure}
\includegraphics[width=\columnwidth,viewport=20 200 575 760,clip=true]{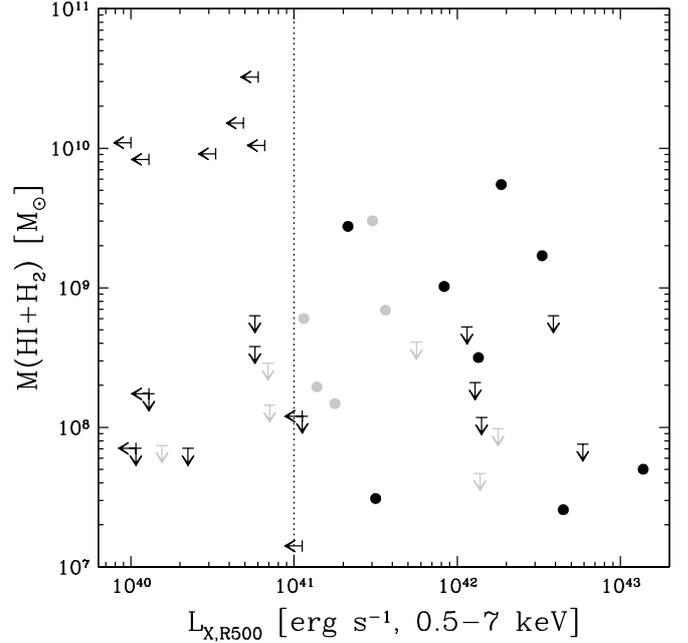}
\caption{\label{fig:McoldLx} Cold gas mass against 0.5-7~keV X-ray luminosity within R$_{500}$ for a subset of our groups. Black points indicate systems in the CLoGS high-richness sample \protect\citep{OSullivanetal17}, for which complete X-ray data are available. Grey points mark additional groups for which X-ray luminosities were drawn from the literature. The dotted line indicates the 10$^{41}$\ergps\ dividing line between X-ray bright and faint systems.}
\end{figure}

Figure~\ref{fig:McoldLx} shows total cold gas mass (\Hi+H$_2$) in the BGEs plotted against the X-ray luminosity of the IGM. Systems in the high-richness subsample of CLoGS, for which complete X-ray data are available, are marked in black, other groups in grey. As expected, the X-ray luminosities drawn from the literature are all detections, but their distribution seems comparable to the high-richness groups, suggesting that their inclusion does not introduce a bias. \citet{OSullivanetal17} divided the CLoGS high-richness groups into X-ray bright systems, with L$_{\rm X,500}$$>$10$^{41}$\ergps and IGM extent >65~kpc, and X-ray faint systems which have lower luminosities, and either small X-ray halos associated with the BGE, or no hot gas at all. There is a large degree of scatter in the plot, and it is clear that the BGEs of X-ray bright groups can possess significant quantities of cold gas. However, the six BGEs with the highest M(\Hi+H$_2$) are all X-ray faint (NGC~924, NGC~940, NGC~1167, NGC~4169, NGC~5084, \object{ESO~507-25}). The two X-ray bright groups with the largest M(\Hi+H$_2$) are NGC~5353 and NGC~5903. The latter has certainly acquired its \Hi\ through tidal stripping \citep{Appletonetal90}, and the former is interacting with, and may have acquired gas from, its neighbor NGC~5354. These results suggest that the presence of an X-ray bright IGM makes it difficult for BGEs to retain very large cold gas reservoirs (\gtsim 10$^{10}$\Msol), even though their cold gas content is not so suppressed as is the case for cluster galaxies. However, since the \Hi\ and X-ray data are not complete, we cannot rule out the possibility that future observations will identify X-ray luminous groups whose BGEs contain $>$10$^{10}$\Msol of cold gas.

\subsubsection{Thermal instability}
For a subset of the X-ray bright groups, we have more detailed information on the thermodynamic state of the IGM. In galaxy clusters, molecular and ionized gas is largely restricted to BCGs with central cooling times $<$1~Gyr, entropies $<$35~keV~cm$^2$, or low minimum values of the ratio of the cooling time to the free fall time in the IGM, min($t_c/t_{ff}$)$\lesssim$25. Low values of min($t_c/t_{ff}$) were considered indicative of the onset of thermal instability in ICM gas, but a number of studies have shown that BCGs have such similar properties that the free-fall time at small radii is almost constant \citep{McNamaraetal16,Hoganetal17,Pulidoetal18}, making cooling time the dominant factor.

In \citet{OSullivanetal17} we examined the relationship between several measures of X-ray cooling (e.g., central temperature profile, central cooling time, entropy at 10~kpc radius) and evidence of recently or currently active radio jets in the BGE. We found that radio jets were closely correlated with declining central temperature profiles and, where it could be calculated, low minimum values of the ratio of the cooling time to the free fall time in the IGM, min($t_c/t_{ff}$)$\lesssim$15. This link with jet activity suggests that at low values of this instability criterion, cooling from the IGM produces sufficient cold material to fuel the AGN \citep[see also][]{ValentiniBrighenti15,Prasadetal15,Prasadetal17,Prasadetal18}. The jet systems also had low central cooling times and entropies, but so did a number of systems without evidence of recent AGN outbursts.

\begin{figure}
\includegraphics[width=\columnwidth,viewport=20 200 575 760,clip=true]{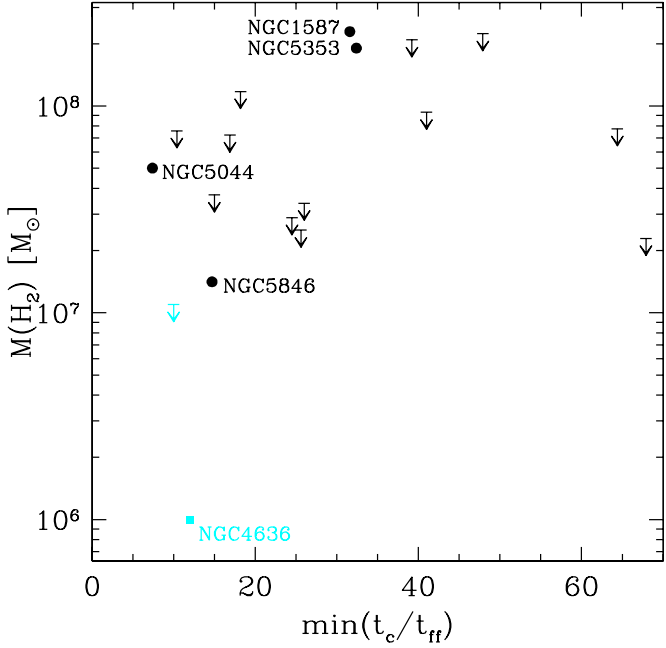}
\caption{\label{fig:TcTff} Molecular gas mass against the thermal instability criterion min(t$_{c}$/t$_{ff}$), drawn from \citet{OSullivanetal17} and \citet{Voitetal15}, for a subset of our group-dominant galaxies. Cyan points indicate non-CLoGS galaxies.}
\end{figure}

Figure~\ref{fig:TcTff} shows molecular gas mass plotted against the thermal instability criterion min($t_c/t_{ff}$) for the subset of systems with sufficient X-ray data to calculate this value. Unfortunately only five of the CO-detected BGEs have high-quality X-ray data available. Three systems with min($t_c/t_{ff}$)$<$15 are detected: NGC~4636, NGC~5044 and NGC~5846. All three have relatively low molecular gas masses (1-50$\times$10$^6$\Msol) but clearly dominate the cool cores of their respective groups, with no near neighbours of comparable size. All three galaxies possess currently or recently active radio jets, as do all the CO-undetected galaxies with min($t_c/t_{ff}$)$\leq$15. The two other CO-detected systems are NGC~1587 and NGC~5353, which have min($t_c/t_{ff}$)$>$30 and M(H$_2$)$\sim$2$\times$10$^8$\Msol, a factor of $\sim$4 greater than NGC~5044, or a factor of $\sim$100 greater than NGC~4636. Both these systems appear to be interacting with neighbouring galaxies, occupy groups without large cool cores, and lack radio jets, though NGC~1587 has diffuse extended radio emission of unknown origin \citep{Giacintuccietal11}. Although it is possible that the molecular gas in these systems is the product of IGM cooling, which past AGN outbursts have been unable to disperse \citep{Prasadetal18}, it seems more plausible that they have acquired their cold gas through tidal interactions.

Comparing molecular gas mass with the IGM entropy at 10~kpc (K$_{10}$, the fixed radius used in cluster studies to avoid resolution biases) we find that for the CLoGS groups, systems with detected molecular gas all have K$_{10}$$\lesssim$35~keV~cm$^2$, but so do seven BGEs without molecular gas. Conversely, two BGEs with active jets but no molecular gas, NGC~1060 and NGC~4261, have K$_{10}$$>$35~keV~cm$^2$. The size of the cooling region in these two groups, based on the downturn of the temperature profile, is $\sim$10-20~kpc, considerably smaller than the typical scale of cool cores in clusters, so it may be that the radius chosen for cluster samples is simply too large to probe the cooling regions of groups.

\begin{figure}
\includegraphics[width=\columnwidth,viewport=20 200 575 760,clip=true]{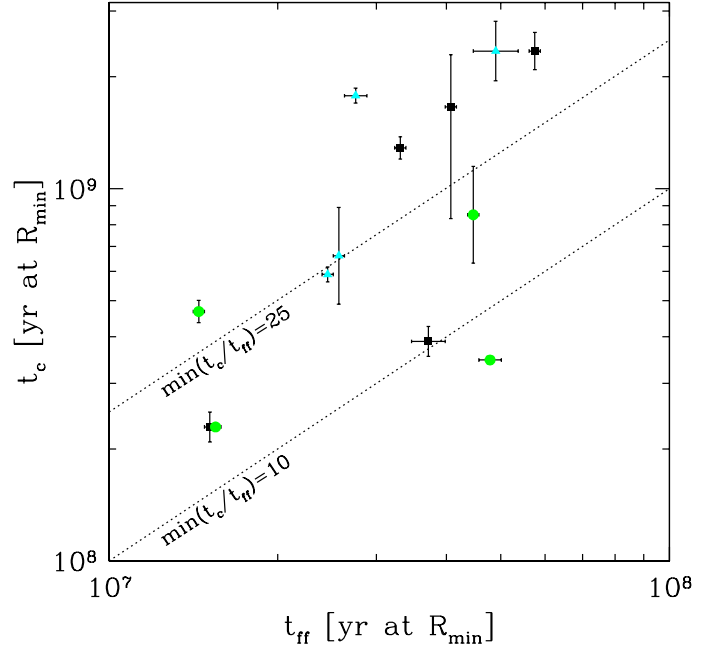}
\caption{\label{fig:tc_vs_tff}cooling time plotted against free-fall time at the radius of minimum $t_c/t_{ff}$ for CLoGS BGEs with sufficient X-ray data. BGEs detected in CO are marked by green circles, those detected only in \Hi\ by cyan triangles, and BGEs with no detected cold gas by black squares Dotted lines show locations of min($t_c/t_{ff}$)=10 and 25.}
\end{figure}

Examining $t_c/t_{ff}$ in more detail we find that, as in clusters, $t_{ff}$ at the radius of minimum $t_c/t_{ff}$ covers a narrow range (10-60~Myr) and is not strongly correlated with min($t_c/t_{ff}$). Cooling times at this radius cover the range $\sim$0.2-3~Gyr. Figure~\ref{fig:tc_vs_tff} shows $t_c$ and $t_{ff}$ at the radius of minimum $t_c/t_{ff}$ plotted against each other for the CLoGS groups, excluding two (NGC~978 and NGC~4008) with very large uncertainties in $t_c$. The four CO-detected sources all have $t_c$$<$1~Gyr, as do two others with \Hi\ detections, and two galaxies with no cold gas. Of the five galaxies with $t_c$$>$1~Gyr, two have \Hi\ detections. We only have information on the distribution of \Hi\ in NGC~5982, which has $t_c$$<$1~Gyr, min($t_c/t_{ff}$)$\sim$25, and so might be expected to be cooling. However, the cold gas is located in a cloud offset from the galaxy center by 6~kpc and 200\kmps\ \citep{Morgantietal06}, perhaps more suggestive of tidal acquisition.

It seems clear from these results that while the connection between cold gas in BGEs and the surrounding IGM is similar to that of BCGs and the ICM, the relationships established in clusters do not appear quite so clearly in groups. Gas morphology suggests one reason for these differences, and we discuss this in the following section.

\section{Discussion}
\label{sec:disc}

Although we have only collected single-dish data for our BGEs, a number of our targets have previously been observed in CO with interferometric arrays and others have \Hi\ imaging or kinematic data, providing information on the distribution of their cold gas content. Table~\ref{tab:morph} lists the information available for our CO-detected systems.

\begin{table*}
\caption{\label{tab:morph}Cold gas morphology of our CO-detected galaxies}
\begin{center}
\begin{tabular}{lclc}
\hline
\hline
Galaxy   & Morphology    & Notes & Reference\\
\hline\\[-3mm]
NGC~128  & Cloud         & CO at velocity of 15~kpc tidally-stripped \Hi\ cloud & 1 \\
NGC~252  & -             & - & - \\
NGC~315  & Clouds        & multiple \Hi\ clouds in absorption and emission & 2 \\
NGC~524  & Disk          & 1.1~kpc CO gas disk & 3 \\
NGC~924  & Disk (probable) & double-horned \Hi\ profile, CO profile has asymmetric peak & 4 \\
NGC~940  & Disk (probable) & double-horned CO and \Hi\ profiles & 5,6 \\
NGC~1106 & -             & CO line profile has asymmetric peak& - \\
NGC~1167 & Disk          & 160~kpc \Hi\ disk & 7 \\
NGC~1587 & -             & - & - \\
NGC~1779 & Disk (probable) & Top hat \Hi\ line profile & 6 \\
NGC~2768 & Disk          & $\sim$1~kpc recently accreted CO polar disk and offset \Hi\ cloud & 8 \\
NGC~2911 & Disk          & 500~pc kinematically misaligned H$_2$ disk, double-horned CO profile & 9 \\
NGC~3665 & Disk          & $\sim$5~kpc CO disk & 10 \\
NGC~4169 & -             & - & - \\
NGC~4636 & Clouds        & Sub-kpc CO clouds, but probable double-peaked \Hi\ profile & 11, 16 \\ 
NGC~5044 & Clouds        & Sub-kpc CO clouds & 12 \\
NGC~5127 & -             & - & - \\
NGC~5353 & Disk          & 1.6~kpc CO disk with broad line profile & 13 \\
NGC~5846 & Clouds        & Sub-kpc CO clouds, but probable double-peaked \Hi\ profile & 11 \\
NGC~7252 & Disk + tidal arms & Post-merger with tidal tails, disk, etc. & 14 \\
NGC~7377 & -             & - & - \\
ESO~507-25 & Disk (probable) & Double-peaked \Hi\ line profile & 15, 17 \\
\hline
\end{tabular}
\end{center}
\tablefoot{References: 
1 \protect\citet{Chungetal12};
2 \protect\citet{Morgantietal09}; 
3 \protect\citet{Crockeretal11};
4 \protect\citet{Springobetal05}; 
5 see paper~I; 
6 from the Extragalactic Distance Database All-Digital \Hi\ catalog, \protect\citet{Courtoisetal09}; 
7 \protect\citet{Struveetal10}; 
8 \protect\citet{Crockeretal08}; 
9 \protect\citet{Muller-Sanchezetal13}
10 \protect\citet{Alataloetal13}; 
11 \protect\citet{Temietal17};
12 \protect\citet{Davidetal14,Davidetal17}; 
13 \protect\citet{Alataloetal15};
14 \protect\citet{Hibbardetal94,Wangetal92};
15 \protect\citet{Doyleetal05};
16 \protect\citet{BottinelliGouguenheim77};
17 \protect\citet{BottinelliGouguenheim79}
}
\end{table*}

Of the 15 systems for which we find some indication of the gas structure, 11 appear to contain gas disks. We include NGC~924, NGC~940, \object{NGC~1779} and ESO~507-25 where no imaging studies have been performed, but CO and/or \Hi\ line profiles suggest rotation, and NGC~7252, where molecular gas is observed in tidal tails as well as a central disk. The molecular disks are generally small, consistent with the limits on star-forming disk size we draw from the radio data. The mass of molecular gas associated with the disks covers a wide range, but includes some of our CO-richest systems, e.g., NGC~940, NGC~3665, NGC~1779, NGC~1167.

In several cases, there is evidence that the cold material is related to tidal interactions or mergers, the prime example being the tidal tails of NGC~7252. NGC~5353 and its close neighbor NGC~5354 are located in the core of the compact group HCG~68, overlap in projection and are only separated by $\sim$250\kmps. We observe CO in emission in NGC~5353 and in absorption in NGC~5354. Previous studies have shown the CO line profile of NGC~5353 to be exceptionally broad. \citet{Alataloetal15} report rotation and a full width at zero intensity linewidth of $\sim$2000\kmps\ in their CARMA data. We measure 655\kmps\ and 700\kmps\ Gaussian FWHM line widths in CO(1-0) and CO(2-1) respectively, in good agreement with the CARMA result. This suggests that a large or highly disturbed disk is present, with NGC~5354 the likely source of the disturbance. Additionally, although we find no information on the morphology of the cold gas in NGC~1587, there is a suggestion that the interaction involves the \Hi\ content of the two galaxies, based on a broad emission linewidth overlapping the velocities of the two stellar components \citep{Gallagheretal81}. The kinematic offsets observed between the gas disks and stars in NGC~2768 \citep{Crockeretal08} and NGC~2911 \citep{Muller-Sanchezetal13} may also indicate that the gas has an external origin, though cooling provides another possible explanation.

Five of our galaxies contain cold gas whose morphology can best be described as clouds. These include the three systems mentioned in Section~\ref{sec:intro}, NGC~4636, NGC~5044, and NGC~5846, which previous studies have shown to have strong similarities to strongly cooling galaxy clusters. ALMA observations show that much of the molecular gas is in sub-kiloparsec-scale clouds with masses comparable to those of Giant Molecular Associations in the Milky Way \citep{Temietal17,Davidetal14,Davidetal17}. These are located within more extended filamentary H$\alpha$ nebulae, and a variety of measurements suggest that both molecular and ionized gas components are the product of cooling. In all three galaxies the cloud velocities are consistent with formation within the galaxy rather than acquisition from tidal stripping, and none of the three has a close gas rich neighbor. However, significant masses of \Hi\ are detected in NGC~4636 and NGC~5846, in both cases with line profiles suggesting the double peak expected from rotation \citep{BottinelliGouguenheim77,BottinelliGouguenheim79}. \Hi\ mapping is needed to determine the location and kinematics of this gas, and its relationship to the ionized and molecular components in the galaxy cores.

The other two galaxies which host cold gas clouds are different. Both contain \Hi\ components whose properties may suggest interactions. In NGC~128 an \Hi\ cloud overlaps one half of the galaxy disk and the near neighbor NGC~127, with a mean velocity between those of the two galaxies. \citet{Chungetal12} suggest that NGC~127 is likely to merge with NGC~128 in the next few hundred Myr, and that the \Hi\ velocity profile indicates that the gas originated in NGC~128. Our CO detection is found at the velocity of the \Hi\ rather than the stellar population. NGC~315 contains multiple \Hi\ structures, including two parsec-scale absorption components and a few 10$^7$\Msol\ cloud seen in emission, offset from the nucleus by a few kiloparsecs. \citet{Morgantietal09} note a similarity in the velocities of the cloud seen in emission and the narrower absorption feature, redshifted by 490\kmps\ compared to the stellar population, and the presence of \Hi\ in neighbouring galaxies, and suggest the most likely origin of the gas is external. There are other examples of offset \Hi\ clouds among our CO-undetected galaxies \citep[e.g., NGC~584 and NGC~5982,][]{Oosterlooetal10,Morgantietal06}, and clear cases of tidal stripping of \Hi\ from group members by the BGE \citep[e.g., NGC~5903,][]{Appletonetal90}.

These differences in gas morphology suggest differences in origin; if NGC~4636, NGC~5044 and NGC~5846 are exemplars of the condensation of cold gas from a cooling IGM, then the gas disks we see in other systems seem likely to arise from some other process. The disturbed morphologies and evidence of interaction we see in some systems suggest that tidal stripping likely plays an important role, with galaxies acquiring gas directly from neighbours, or potentially from intergalactic gas structures left behind prior galaxy encounters.

Although we only detect CO in three, eight of our galaxies have measured min($t_c/t_{ff}$)$<$20, suggesting that they are likely sites for gas condensation. These include NGC~5813, where the presence of an H$\alpha$ nebula, [C\textsc{ii}] emission, active radio jets and multiple cavity pairs strongly suggest cooling has occurred \citep{Randalletal11,Werneretal14,Randalletal15}. Our deep IRAM 30m observation puts a relatively strong upper limit on the molecular gas mass, M(H$_2$)$<$1.1$\times$10$^7$\Msol, but this is still a factor $\sim$10 greater than the detected molecular gas mass in NGC~4636. The limits on the other systems are even higher, exceeding the detected  mass in NGC~5846 and in most cases NGC~5044. The sensitivity of our observations may be simply insufficient to trace the products of IGM cooling in some galaxy groups. If cooling in a moderately luminous group such as NGC~4636 only produces a few 10$^6$\Msol\ of molecular gas, we will need significantly deeper observations to study the process of IGM cooling and condensation in groups. 

It is also likely that we are observing the eight groups at different stages of the feedback cycle. Simulations suggest that the cycle of cooling and feedback can be irregular, that individual heating episodes may not increase min($t_c/t_{ff}$) by large factors, and that AGN outbursts may not disperse all the molecular gas \citep{Prasadetal18}. If heating episodes are short compared to the cooling phase, we might expect to see a range of molecular gas masses, even in systems with relatively low min($t_c/t_{ff}$), since most would be in the process of building up their cold gas reservoir. NGC~5044 and NGC~4636 may provide an example: NGC~5044 has the higher M(H$_2$) and appears to be in the very early stages of an AGN outburst \citep{Davidetal17}, whereas NGC~4636 may have recently completed an active phase, leaving only a little molecular gas. We can therefore speculate that some of the CO-undetected systems with low min($t_c/t_{ff}$) have low molecular gas masses because they are still recovering from past outbursts. Again, deeper CO observations would be needed to explore this possibility.

The other indicators of cooling identified in galaxy clusters, cooling times $<$1~Gyr and K$_{10}$$<$35~keV~cm$^2$ suggest that we might expect to find cooled material in an even larger fraction of our BGEs. However, some of the BGEs meeting these criteria contain cold gas which looks more likely to have been acquired from other galaxies. While it is possible that in some groups both ICM cooling and accretion from other galaxies contribute to the gas reservoir, we must question whether the cooling thresholds established in galaxy clusters are appropriate for groups. \citet{OSullivanetal17} argued that central cooling time thresholds determined from clusters tended to misclassify groups; almost all groups end up classed as strong cool cores, since group cooling times are systematically shorter than clusters. This may indicate that the thresholds for the onset of thermal instability scale with IGM temperature. Further investigation with samples containing statistically significant numbers of groups and clusters is needed to resolve this question.

Although the relatively small molecular gas masses we observe in systems like NGC~4636 and NGC~5846 are difficult to detect, they are sufficient to power the radio AGN we observe. As discussed in paper~I, if accretion of molecular gas is only 1\% efficient, molecular gas masses of a few 10$^6$-10$^7$\Msol\ are sufficient to produce outbursts with total power up to $\sim$10$^{59}$~erg, comparable to the most luminous radio galaxies in our sample. Molecular gas is therefore a plausible fuel source for the AGN we observe, even in those systems where we fail to detect CO.

However, this does not imply that IGM cooling is always the driver for AGN outbursts. NGC~315 and NGC~1167 are among our most powerful radio sources, are detected in CO, but both show signs of having acquired their cold gas component through interactions, and indeed NGC~1167 is located in a group without an X-ray detected IGM. This suggests that AGN outbursts can be triggered by galaxy interactions, fuelled either by the disturbance of stable gas structures or by an influx of gas from the other galaxy, and that such outbursts may be relatively common. Given the radio power of some of these sources, it seems likely that their impact on the group environment may be profound. Simple models of cooling in X-ray bright groups, regulated by mild near-continuous AGN feedback, may need to include disruptions from occasional powerful outbursts whose timing is unrelated to the thermodynamic state of the IGM.

In addition to IGM cooling and acquisition of gas through tidal stripping, it seems likely that our sample contains some systems whose cold component arises from stellar mass loss within the BGE \citep{PadovaniMatteucci93,CiottiOstriker07,ParrottBregman08,Ciottietal15,Negrietal15}. We certainly observe galaxies with molecular disks whose radio luminosity is consistent with the star formation rates observed in the relaxed disks of spiral galaxies. The relative importance of these different sources of cold gas is unclear, but further interferometric observations could offer insight. If, as has been suggested \citep{Davisetal11}, gas produced by the stellar mass loss should form kiloparsec-scale disks aligned with the stellar population, then imaging and velocity mapping of our CO-detected BGEs will allow us to determine the fraction of systems in each category, and the range of molecular gas masses that can be built up via each method. Supporting observations of atomic and ionized gas would provide independent confirmation and additional understanding of the different gas phases in these objects. As we work to complete the X-ray coverage and radio analysis of CLoGS, such observations offer an opportunity to understand the various factors affecting the development of groups and their dominant galaxies.

\section{Summary and Conclusions}
\label{sec:conc}

We have presented 33 new IRAM~30m and APEX observations and results from 3 APEX archival observations, using CO(1-0) and CO(2-1) measurements to estimate or put upper limits on the molecular gas content of 36 group-dominant early-type galaxies. We report 13 detections, 12 of CO emission arising from the target galaxy or associated gas structures, as well as the detection of CO in absorption in NGC~5354. These observations complete our survey of the dominant galaxies of the Complete Local-volume Group Sample (CLoGS).

We find that 21 of the 53 CLoGS BGEs are detected in CO, suggesting that group-dominant galaxies may be more likely to contain molecular gas than the general population of early-type galaxies. Our observations are typically sensitive to molecular gas masses of a few 10$^7$\Msol\, and across the whole CLoGS sample we detect galaxies containing 1-610$\times$10$^7$\Msol. The majority of our galaxies have low star formation rates (0.01-1\Msolpyr) but short depletion times ($<$1~Gyr) indicating that, averaged over the population, their molecular gas reservoirs must be replenished on relatively rapid timescales. Using \Hi\ mass measurements from the literature, we find that at least 27 of the 53 CLoGS BGEs contain \Hi. This is similar to the fraction of nearby early-type galaxies outside the Virgo cluster which contain \Hi, but significantly greater than the fraction of cluster members with \Hi. The \Hi\ masses in our galaxies fall in the range of a few 10$^6$ to a few 10$^{10}$\Msol, with typical M(H$_2$)/M(\Hi) in the range 0.01-1. Averaging over all galaxies where gas is detected, the mean molecular and atomic gas content of the BGEs is 1\% and 5\% of the stellar mass, respectively.

Comparing 1.4~GHz radio luminosity with molecular gas content, we find that AGN likely provide the dominant radio contribution in most of our galaxies, but that the predicted radio luminosity from star formation forms a lower boundary to the distribution, suggesting that in a subset of our galaxies the molecular gas is forming stars much as in spiral galaxies. A significantly higher fraction (60$\pm$16\%) of our BGEs have radio luminosities more than a factor of 5 above the predicted M(H$_2$):L$_{\rm 1.4~GHz}$ relation than is found for the general population of early-type galaxies (9$\pm$4\%), confirming that group-dominant galaxies are more likely to host radio luminous AGN. However, we see no clear connection between molecular gas mass and radio luminosity, and roughly half of the most radio luminous sources in our sample (L$_{\rm 1.4~GHz}$$>$5$\times$10$^{24}$~W~Hz$^{-1}$) are undetected in CO.

Our data hint at an anti-correlation between the richness of the group and the presence of molecular gas in the dominant galaxy, with 20/42 groups with $R$=2-5 having a BGE with detected CO, compared to only 1/11 with $R$=6-8 (a 2.75$\sigma$ significant difference). We see no clear connection between the presence of a detected X-ray bright IGM and CO in the dominant galaxy, but our X-ray coverage of the sample is as yet incomplete. We do find that the most \Hi-rich BGEs are located in X-ray faint groups, but additional \Hi\ and X-ray observations are needed to confirm this result.

Considering the information in the literature on the cold gas distribution in our targets, we find that many of our CO-detected galaxies contain gas disks. A number of systems show indications of ongoing or past tidal interactions, suggesting that they may have acquired material from gas-rich neighbours. These systems include some of our most radio luminous AGN, suggesting that mergers and interactions may be the source of some powerful AGN outbursts. We find that the three galaxies NGC~4636, NGC~5846 and NGC~5044, which previous works have identified as likely to host cooling flows from which their molecular gas has condensed, are distinct from the other galaxies for which we have morphological data, containing small-scale molecular gas clouds with low relative velocities rather than disks or tidal structures. Although our single-dish data reveal 2-7 times more CO(2-1) emission in NGC~4636 and NGC~5846 than previous ALMA observations, they are still amongst the CO-poorest galaxies detected in our sample, raising the possibility that IGM cooling produces only relatively small molecular gas masses in galaxy groups. Further deep observations to measure the molecular gas content of currently undetected galaxies would help to answer the question of the effectiveness of IGM cooling in groups. Interferometric mapping is also needed to determine the relative importance of IGM cooling, tidal interactions and stellar mass loss in group-dominant galaxies.

%%%%%%%%%%%%%%%%%%%%%%%% acknowledgments
\begin{acknowledgements}
  The IRAM and APEX staff are gratefully acknowledged for their help in the
  data acquisition. We thank the anonymous referee for their constructive
  and helpful comments, which have materially improved the paper.  F.C.
  acknowledges the European Research Council for the Advanced Grant Program
  Number 267399-Momentum. P.S. acknowledges acknowledges support from the
  ANR grant LYRICS (ANR-16-CE31-0011). E.O'S.  acknowledges support from
  the National Aeronautics and Space Administration through Chandra Awards
  Number GO6-17121X and GO6-17122X issued by the Chandra X-ray Observatory
  Center, and thanks M.~Gitti and K.~Kolokythas for their comments on the
  draft. A.B. acknowledges support from NSERC through the Discovery Grants
  program. J.L. acknowledges support from the Research Grants Council of
  Hong Kong through grant 17303414 for the conduct of this work.This work
  made use of the NASA/IPAC Extragalactic Database (NED), and of the
  HyperLeda database.
\end{acknowledgements}
%%%%%%%%%%%%%%%%%%%%%%%%%%%%%%%%%%%%%

\bibliographystyle{aa}
\bibliography{../paper}

\begin{appendix}
\section{IRAM and APEX beam sizes}
\label{opt}
\begin{sidewaysfigure*}
\hspace{-1.5cm}
\includegraphics[height=28cm,angle=-90]{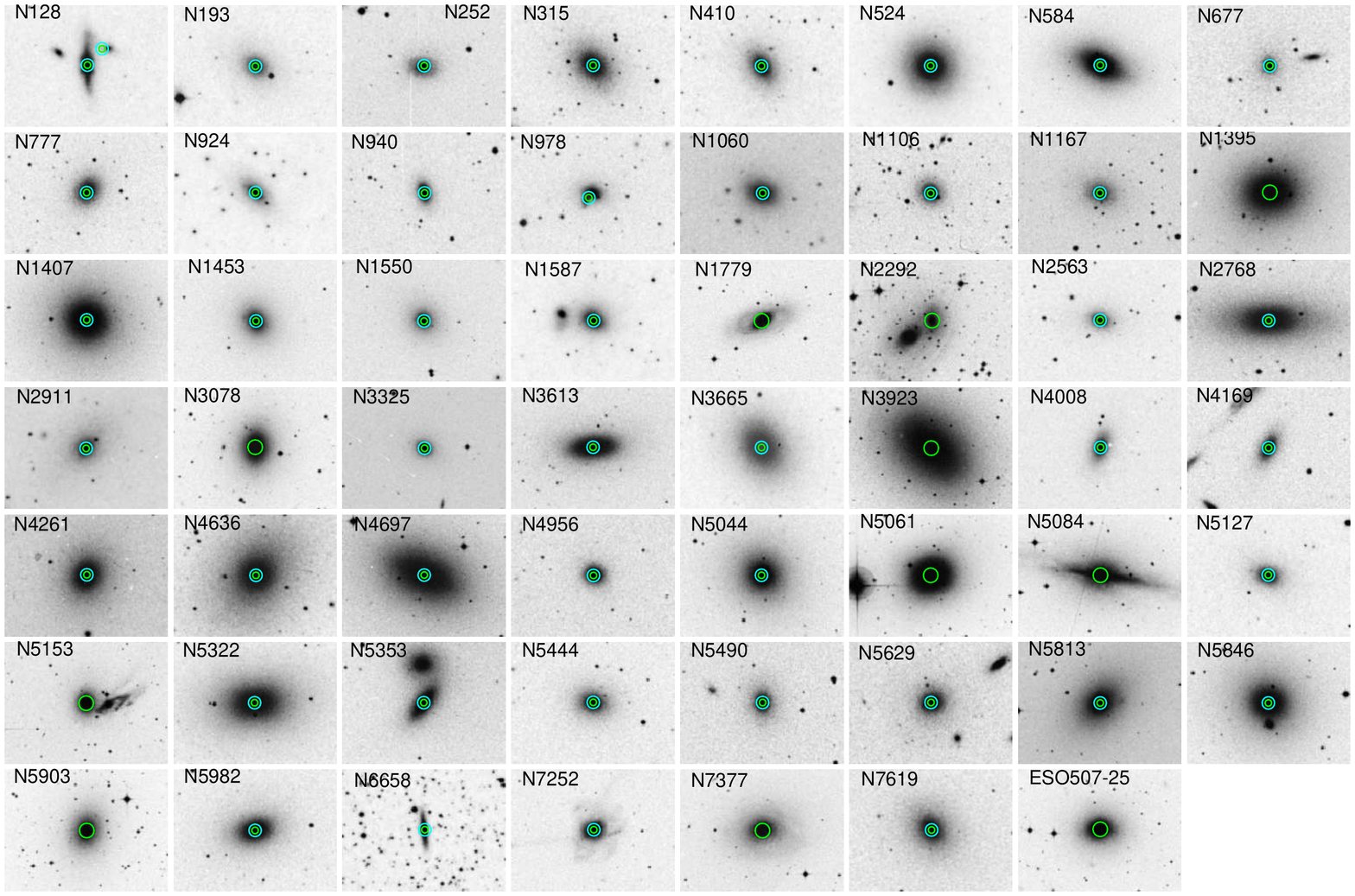}
\vspace{-3cm}
\caption{\label{fig:beams}DSS optical images of the BGEs with CO beams overlaid, CO(1-0) in cyan, CO(2-1) in green.}
\end{sidewaysfigure*}

Figure~\ref{fig:beams} shows Digitized Sky Survey (DSS) optical images of the 55 BGEs for which we have CO data, with IRAM and APEX beams marked. Typical beam sizes are 23\arcs\ and 12\arcs\ for the IRAM~30m CO(1-0) and CO(2-1) respectively, and 27\arcs\ for the APEX CO(2-1) observations.

\section{Combined measurements for the full sample}
\label{bigtab}
Table~\ref{tab:bigtab} lists key parameters for the full CLoGS sample, plus NGC~4636 and NGC~5813, including systems from paper~I whose values have been updated.

\begin{table*}
\caption{\label{tab:bigtab}Key data for the full sample}
\vspace{-6mm}
\begin{center}
\begin{tabular}{lccccclc}
\hline\hline
Galaxy & z & D$_L$ & log M(HI) & log L$_{\rm FIR}$ & M(H$_2$) & F$_{1.4GHz}$ & H\textsc{i} Reference\\
       &   & [Mpc] & [M$_\odot$] & [L$_\odot$] & [10$^8$ M$_\odot$] & [mJy] & \\
\hline
NGC128    & 0.014146 & 58.54 & 8.65    & 9.84$\pm$0.03  & 1.76$\pm$1.5  & 1.5   & 1 \\
NGC193    & 0.014723 & 63.80 & -       & -              & $<$1.35       & 1710  & - \\
NGC252    & 0.016471 & 71.65 & 9.56    & 9.84$\pm$0.05  & 6.29$\pm$0.79 & 2.5   & 2 \\
NGC315    & 0.016485 & 71.60 & 7.86    & 9.56$\pm$0.06  & 0.74$\pm$0.17 & 6630  & 3 \\
NGC410    & 0.017659 & 76.67 & -       & -              & $<$1.17       & 5.8   & - \\
NGC524    & 0.008016 & 34.60 & 6.76    & 9.39$\pm$0.06  & 1.9$\pm$0.23  & 3     & 4 \\
NGC584    & 0.006011 & 18.71 & 8.08    & 8.15$\pm$0.06  & $<$0.11       & 0.6   & 5 \\
NGC677    & 0.017012 & 73.90 & 9.23    & -              & $<$2.25       & 21    & 2 \\
NGC777    & 0.016728 & 72.60 & -       & 7.91$\pm$0.18  & $<$2.08       & 7     & - \\
NGC924    & 0.014880 & 63.61 & 9.96    & 9.50$\pm$0.23  & 0.52$\pm$0.10 & 0.9   & 6 \\
NGC940    & 0.017075 & 74.20 & 9.33$^*$    & 10.06$\pm$0.05 & 61.0$\pm$2.15 & 8     & 7 \\
NGC978    & 0.015794 & 68.66 & -       & 8.96$\pm$0.08  & $<$0.70       & 0.7   & - \\
NGC1060   & 0.017312 & 75.20 & -       & 10.51$\pm$0.03 & $<$0.76       & 9     & - \\
NGC1106   & 0.014467 & 63.26 & -       & 10.03$\pm$0.06 & 6.81$\pm$0.33 & 132   & - \\
NGC1167   & 0.016495 & 71.60 & 10.01   & 9.59$\pm$0.07  & 3.3$\pm$0.28  & 1840  & 8 \\
NGC1395   & 0.005727 & 22.39 & $<$8.45 & 8.12$\pm$0.08  & $<$0.27       & 1.1   & 9 \\
NGC1407   & 0.005934 & 25.50 & $<$8.57 & 8.49$\pm$0.07  & $<$0.34       & 89    & 9 \\
NGC1453   & 0.012962 & 53.54 & 9.01    & 9.10$\pm$0.06  & $<$0.78       & 28    & 10 \\
NGC1550   & 0.012389 & 52.13 & -       & -              & $<$0.47       & 17    & - \\
NGC1587   & 0.012322 & 53.30 & 9.40    & -              & 2.3$\pm$0.48  & 131   & 11 \\
NGC1779   & 0.011051 & 44.44 & 9.26    & 9.73$\pm$0.08  & 4.57$\pm$0.60 & 5.4   & 7 \\
NGC2292   & 0.006791 & 29.57 & 9.35    & 9.26$\pm$0.03  & $<$0.47       & -     & 12 \\
NGC2563   & 0.014944 & 64.39 & $<$8.73 & -              & $<$0.93       & 0.7   & 13 \\
NGC2768   & 0.004580 & 19.70 & 7.81    & 8.66$\pm$0.05  & 0.18$\pm$0.01 & 15    & 14 \\
NGC2911   & 0.010617 & 45.84 & 9.31    & 8.83$\pm$0.07  & 2.66$\pm$0.31 & 56    & 2 \\
NGC3078   & 0.008606 & 32.66 & $<$8.78 & -              & $<$0.23       & 310   & 9 \\
NGC3325   & 0.018873 & 79.67 & 9.18    & 4.10$\pm$0.01  & $<$1.57       & -     & 15 \\
NGC3613   & 0.006841 & 29.50 & $<$7.32 & -              & $<$0.46       & $<$0.3 & 16 \\
NGC3665   & 0.006901 & 29.70 & $<$7.34 & 9.74$\pm$0.05  & 6.0$\pm$0.32  & 133   & 16 \\
NGC3923   & 0.005801 & 21.28 & $<$8.41 & 7.38$\pm$0.08  & $<$0.29       & 1.0   & 9  \\
NGC4008   & 0.012075 & 53.96 & $<$8.49 & 8.87$\pm$0.20  & $<$0.73       & 11    & 13 \\
NGC4169   & 0.012622 & 56.67 & 10.03   & 10.46$\pm$0.01 & 1.44$\pm$0.34 & 1.1   & 13 \\
NGC4261   & 0.007378 & 29.38 & $<$8.69 & 8.30$\pm$0.11  & $<$0.37       & 19700 & 9  \\
NGC4636   & 0.003129 & 13.61 & 8.84$^*$    & 7.94$\pm$0.03  & 0.010$\pm$0.003 & 78  & 17 \\
NGC4697   & 0.004140 & 17.80 & $<$8.14 & 8.63$\pm$0.02  & $<$0.07       & 0.4   & 2 \\
NGC4956   & 0.015844 & 70.90 & 9.05    & 9.54$\pm$0.09  & $<$0.75       & -     & 9 \\
NGC5044   & 0.009280 & 40.10 & $<$8.96 & 8.66$\pm$0.15  & 0.5$\pm$0.21  & 36    & 9 \\
NGC5061   & 0.006945 & 28.26 & $<$8.66 & 7.40$\pm$0.10  & $<$0.43       & $<$1.0 & 9 \\
NGC5084   & 0.005741 & 24.09 & 10.18   & 8.87$\pm$0.05  & $<$0.16       & 46    & 18 \\
NGC5127   & 0.016218 & 70.40 & 9.68    & -              & 0.77$\pm$0.05 & 1980  & 2 \\
NGC5153   & 0.014413 & 60.54 & -       & -              & $<$1.75       & -     & - \\
NGC5322   & 0.005937 & 25.60 & $<$7.19 & 8.84$\pm$0.05  & $<$0.58       & 78    & 16 \\
NGC5353   & 0.007755 & 35.43 & 9.72    & 8.74$\pm$0.04  & 1.90$\pm$0.25 & 40    & 13 \\
NGC5444   & 0.013169 & 60.46 & $<$8.79 & -              & $<$0.64       & 660   & 13 \\
NGC5490   & 0.016195 & 71.94 & -       & -              & $<$0.64       & 1300  & - \\
NGC5629   & 0.015004 & 67.73 & -       & -              & $<$0.91       & 4.6   & - \\
NGC5813   & 0.006525 & 29.92 & $<$7.94 & 8.09$\pm$0.06  & $<$0.11       & 15    & 9 \\
NGC5846   & 0.005711 & 23.12 & 8.48$^*$    & 8.05$\pm$0.06  & 0.14$\pm$0.06 & 21    & 19 \\
NGC5903   & 0.008556 & 31.48 & 9.48    & -              & $<$0.58       & 320   & 20 \\
NGC5982   & 0.010064 & 43.50 & 7.49    & 7.87$\pm$0.05  & $<$0.25       & $<$0.5 & 14 \\
NGC6658   & 0.014243 & 63.97 & -       & -              & $<$0.71       & -     & - \\
NGC7252   & 0.015984 & 69.40 & 9.61    & 10.67$\pm$0.03 & 58.0$\pm$8.70 & 25    & 21 \\
NGC7377   & 0.011138 & 46.73 & -       & 9.47$\pm$0.05  & 4.74$\pm$0.44 & 3.1   & - \\
NGC7619   & 0.012549 & 54.30 & 7.41$^*$    & 9.17$\pm$0.12  & $<$0.29       & 20    & 22 \\
ESO507-25 & 0.010788 & 45.21 & 10.50   & 9.46$\pm$0.05  & 4.23$\pm$0.56 & 24    & 23 \\
\hline
\end{tabular}
\end{center}
\vspace{-5mm}
\tablefoot{H\textsc{i} masses marked * may be unreliable, see text for discussion of conflicting measurements.  
References: 
1 \protect\citet{Chungetal12};
2 from the Arecibo Legacy Fast ALFA Survey (ALFALFA), \protect\citet{Haynesetal11,Haynesetal18};
3 \protect\citet{Morgantietal09}; 
4 \protect\citet{Oosterlooetal10};
5 \protect\citet{Haynesetal90};
6 \protect\citet{Springobetal05};
7 from the Extragalactic Distance Database All-Digital \Hi\ catalog, \protect\citet{Courtoisetal09}; 
8 \protect\citet{Struveetal10};
9 \protect\citet{SerraOosterloo10};
10 \protect\citet{Huchtmeier94};
11 \protect\citet{Gallagheretal81};
12 \protect\citet{Barnesetal99};
13 \protect\citet{HuchtmeierRichter89};
14 \protect\citet{Morgantietal06};
15 from HyperLEDA \protect\citet{Patureletal03b};
16 \protect\citet{Serraetal12};
17 \protect\citet{Knappetal78,BottinelliGouguenheim77};
18 from the H\textsc{i} Parkes All-Sky Survey (HIPASS), \protect\citet{Koribalskietal04}
19 \protect\citet{BottinelliGouguenheim79}
20 \protect\citet{Appletonetal90};
21 \protect\citet{HibbardvanGorkom96};
22 \protect\citet{Serraetal08};
23 \protect\citet{Doyleetal05}
}
\end{table*}  

\end{appendix}

\end{document}